\documentclass[12pt]{iopart}

\expandafter\let\csname equation*\endcsname\relax
\expandafter\let\csname endequation*\endcsname\relax
  
\usepackage{graphicx}
\usepackage{amsmath}
\usepackage{bm}
\usepackage{latexsym} 
\usepackage{amstext}
\usepackage{amsxtra}
\usepackage{float}
\usepackage[colorlinks]{hyperref}
\usepackage{times} 
\usepackage{ulem}
\usepackage{amssymb}

\begin{document}

\newcommand{\ket}[1]{ |#1  \rangle}
\newcommand{\bra}[1]{  \langle #1 |}

\title{Dissipative cooling of spin chains by a bath of dipolar particles}
\author{M. Robert-de-Saint-Vincent$^{1,2}$, P. Pedri$^{2,1}$, and B. Laburthe-Tolra$^{1,2}$}
\address{$^1$ CNRS, UMR 7538, Laboratoire de Physique des Lasers, F-93430 Villetaneuse, France \\
$^2$ Universit\'e Paris 13, Sorbonne Paris Cit\'e, LPL, F-93430 Villetaneuse, France}

\ead{martin.rdsv@univ-paris13.fr}
\vspace{10pt}
\begin{indented}
\item[]\today
\end{indented}

\begin{abstract}
We consider a spin chain of fermionic atoms in an optical lattice, interacting with each other by super-exchange interactions. We theoretically investigate the dissipative evolution of the spin chain when it is coupled by magnetic dipole-dipole interaction to a bath consisting of atoms with a strong magnetic moment. Dipolar interactions with the bath allow for a dynamical evolution of the collective spin of the spin chain. Starting from an uncorrelated thermal sample, we demonstrate that the dissipative cooling produces highly entangled low energy spin states of the chain in a timescale of a few seconds. In practice, the lowest energy singlet state driven by super-exchange interactions is efficiently produced. This dissipative approach is a promising alternative to cool spin-full atoms in spin-independent lattices. It provides direct thermalization of the spin degrees of freedom, while traditional approaches are plagued by the inherently long timescale associated to the necessary spatial redistribution of spins under the effect of super-exchange interactions.   
\end{abstract}

\date{\today}

\maketitle

\section{Introduction}

The coupling of quantum systems to environments can lead to various consequences: typically it produces decoherence towards classical behaviors \cite{Zurek2003}, but specific situations are in the spotlight that would on the contrary produce quantum correlations \cite{Benatti2003, Diehl2008, Verstraete2009}. This is an exciting new paradigm, and also opens the fascinating perspective of "environment engineering" to protect or produce entangled states, for quantum information processing and for quantum simulation. Subjecting the various many-body systems explored with cold atoms to well-controlled dissipative processes is a promising prospect to explore the diversity of these situations \cite{Barontini2013, Labouvie2016}. Among the many-body models implemented on ultracold atoms and molecules, lattice spin models have attracted a large attention \cite{dePaz2013, dePaz2016, Martin2013, Yan2013, Greif2013, Barredo2015, Labuhn2016, Hart2015, Cheuk2016, Mazurenko2017, Zeiher2016, Schempp2015}. The question of how dissipation affects them and whether it can be tuned to specific purposes emerges spontaneouly in the context of excited Rydberg atoms \cite{Lee2011, Gunter2013, Hoening2014, Schonleber2017}. It also appears to be an important prospect for ground state atoms in optical lattices \cite{Diehl2008, Yi2012, Bardyn2013, Goto2017,  Kantian2018, Kaczmarczyk2016, Griessner2006, Fischer2016, Wolf2014}, and starts to be experimentally explored \cite{Labouvie2016}.

Recent proposals \cite{Diehl2010, Yi2012, Bardyn2013, Kaczmarczyk2016} involve the use of light as a bath, and spontaneous emission as the dissipative  process. In the present work, we explore binary atomic mixtures \cite{Diehl2008, Griessner2006, Schmidt2018}, one species acting as many-body quantum system, the other as bath. Namely, the quantum system is a spin chain of fermionic atoms, and it is coupled to a Bose Einstein condensate of a different species. The low-energy many-body states of the spin chain are driven by nearest-neighbour super-exchange interactions.  
Magnetic dipole interaction between fermions and bath lead to spin flips in the chain, associated with spontaneous phonon emission in the BEC. Thus, spin-thermalization can arise, due to the spin-orbit coupling conveyed by dipole-dipole interactions, an effect which is connected to the Einstein-de Haas effect \cite{Kawaguchi2006}. As we show here, spin-orbit coupling offers a possibility to directly cool the \textit{collective} spin degrees of freedom in a spin-chain. Such a collective coupling of the spin chain to phonons in the BEC can be seen as an analog of superradience in quantum optics, with the spontaneous collective emission of phonons instead of photons.
We point out that the possibility to couple to the total spin of the chain is particularly relevant to the quantum simulation of the spin-full lattice Hubbard model with ultra-cold atoms. Indeed, in most experiments the collective spin is a conserved quantity which is typically not under control. This can become a limitation to reach the many-body ground state which has a singlet character at half-filling \cite{Auerbach2012}. Our work suggests for the first time to use the intrinsic spin-orbit coupling provided by dipolar interactions, in order to dynamically couple to the collective spin length of a fermionic spin chain and allow reaching its many-body ground state. This approach is complementary to previous studies which used dipolar interactions to cool the mechanical degrees of freedom of a thermal or quantum gas taking benefit of their spin degrees of freedom\,\cite{Fattori2006, Naylor2015B}.

Formally, as presented in sections 2 and 3, we consider a fermionic gas with two spin states in an optical lattice. Due to repulsive contact interactions a Mott insulating regime is reached, in which density fluctuations are strongly reduced \cite{Jordens2008, Schneider2008}. At half filling, the system then consists of a spin chain, described by the Heisenberg spin-spin Hamiltonian. 
To complement this spin-chain Hamiltonian, we consider dipole-dipole interactions between atoms in the spin chain and a spatially overlapping BEC. 
We treat the interaction with this bath using the Fermi golden rule approach. For quantitative predictions, we consider in section 4 a 1D spin chain made of alkali atoms (for which inter-atomic dipole-dipole interactions can safely be neglected compared to super-exchange energies), which interact via dipole-dipole interactions with a bath of a highly dipolar species such as Dy, Er, or Cr \cite{Lu2011, Lu2012, Aikawa2012, Aikawa2014, Griesmaier2005, Naylor2015A}. We find that for realistic experimental parameters, the spin-chain thermalizes close to its highly entangled (singlet) ground state in a few seconds. The final spin-chain temperature is close to the initial temperature of the BEC. Our simulations indicate that the thermalization rate is roughly independent of the length of the spin chain.

\begin{figure}
  \begin{center}
      \includegraphics[width=0.5\columnwidth]{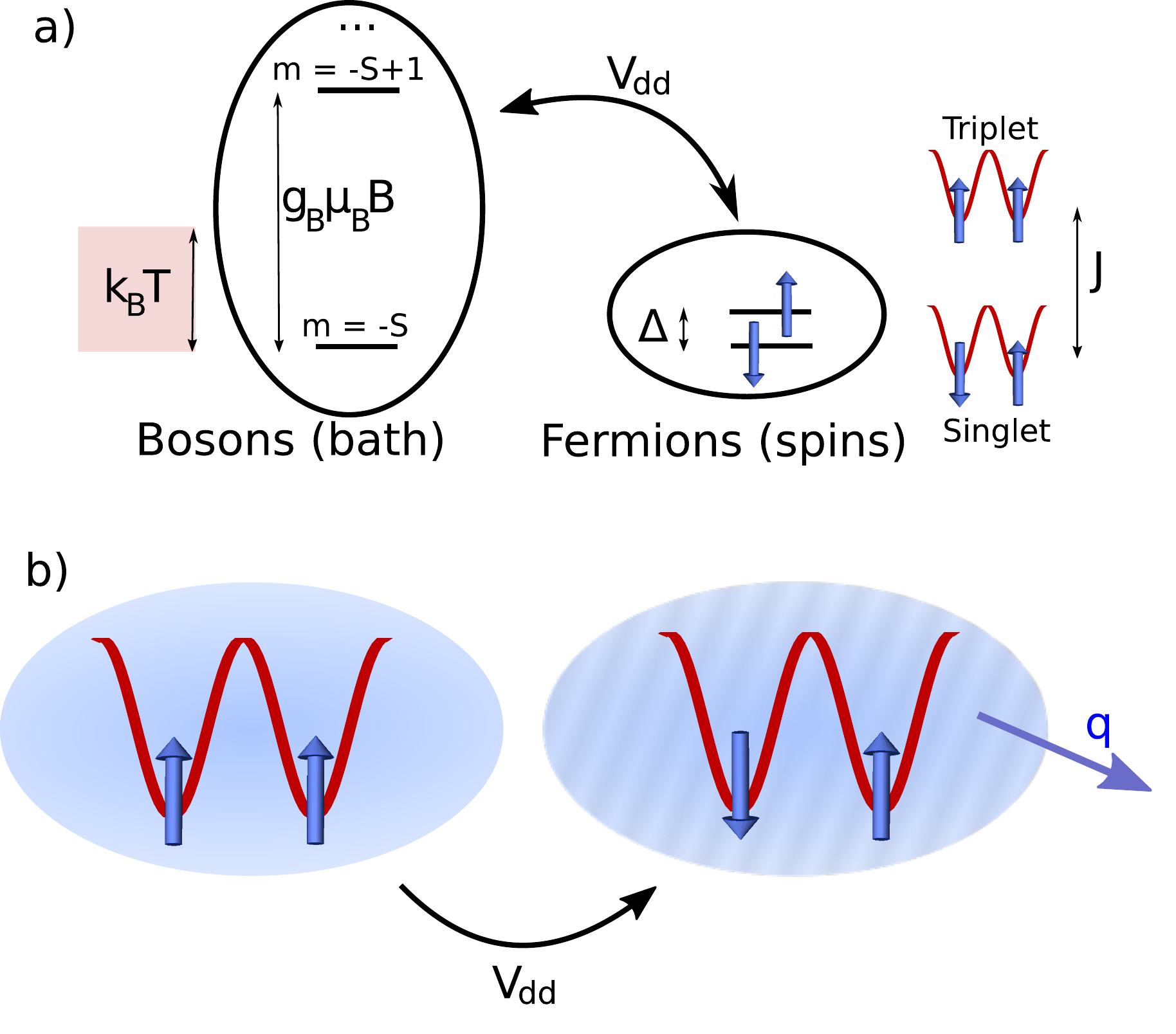}
      \caption{{\bf Spin cooling mechanism.} a) The spin chain is composed of fermions, each of which has two quasi-degenerate spin states. $\Delta$ is the difference of single-particle energy between the two fermionic spin states.
      They interact via dipolar interactions with a bath of bosons with a strong magnetic dipole moment, polarized in a magnetic field. Provided $\Delta \ll k_B T \ll g_B \mu_B B$, inter-species dipolar interactions free the magnetization of the fermions, but energy conservation maintains the magnetization of the bosons. 
      b) The dipolar interaction enables changes in the collective spin chain state, such as changing neighbour correlations from triplet to singlet. For a chain of fermions with spin super-exchange interaction energies $\sim J > k_B T$, the spin chain can lower its energy by exciting a density mode of the bath. \label{fig:overview}}
  \end{center}
\end{figure}

\section{General idea}
\label{section:numerics}
This work focuses on 1D chains of interacting spinful fermionic atoms, prepared in optical lattices. We consider a system with effective spin 1/2 particles - that can be realized by energetically selecting two Zeeman states of an alkali atom in its ground state. These fermions occupy the ground Bloch band of the lattice. Dipole-dipole interactions between fermions are assumed to be negligible.
Fermionic Mott insulators are observed when the on-site interaction $U$ between spin states dominates as compared to the Bloch band width, proportional to the hopping term $t$, and when the temperature $k_B T \ll U$ \cite{Rigol2003, DeLeo2008, Jordens2008, Schneider2008,  Greif2016}. %Cheuk2016
There, for one fermion per site the evolution is well depicted by an effective next-neighbor magnetic interaction Hamiltonian between localized spins \cite{Levy2013}. In 1D,
\begin{equation}
H_{mag} = J  \sum_{site : m} \vec \Sigma_{eff}(m) \cdot  \vec \Sigma_{eff}(m+1)
\end{equation}
with $J = (-U + \sqrt{16 t^2 + U^2})/2 \simeq  4 t^2/U$. This involves effective spin half operators $\vec \Sigma_{eff}$, that do not generically coincide with the true spin operators $\vec \Sigma$ of the atom. For example, if the fermionic species is $^{40}K$, its true spin length is $9/2$. Two selected states could be $\ket{m_F = -9/2}, \ket{m_F = -7/2}$. $\Sigma_{eff}$ are the Pauli spin-1/2 matrices, for example $\Sigma_{eff}^z\ket{-9/2} = -\frac{1}{2} \ket{-9/2}$ while $\Sigma^z\ket{-9/2} = -\frac{9}{2} \ket{-9/2}$. 

In this paper we will describe 1D spin chains with lengths up to 7 sites. The eigenstates are generically strongly entangled, favoring next-neighbor singlet correlations at low energy and triplet correlations at high energy.
For example, if we consider only two sites, the singlet state $\ket{S}$ is the ground state, with energy $-3J/4$. The three triplet states  $\ket{T_0}, \ket{T_{+1}},\ket{T_{-1}}$ with pseudo-spin projection $0, \pm 1$, share the same energy $J/4$. 

To open up the dissipative evolution between these spin eigenstates, a bath of strongly dipolar bosons is introduced (such as Dy, Er, Cr), which interacts with the fermion spins. Dipolar interactions between one bosonic atom spin $\vec S$ at position $\vec r$ and one fermionic atom spin $\vec \Sigma$ at position $\vec r + \vec{dr}$, with respective Land\'e factors $g_B$ and $g_F$, are anisotropic:
\begin{equation}
V_{dd}(\vec r, \vec r + \vec{dr}) = \frac{\mu_0}{4 \pi} \frac{g_B g_F \mu_B^2}{\vert \vec{dr} \vert^5} \left( \vert \vec{dr} \vert^2 \vec \Sigma \cdot \vec S - 3 (\vec \Sigma \cdot \vec{dr}) (\vec S \cdot \vec{dr}) \right)
\label{eq:Vddr}
\end{equation}
using dimensionless (i.e., divided by $\hbar$) spin operators. Here $\mu_0$ is the vacuum magnetic permeability and $\mu_B$ the Bohr magneton. This interaction involves the \textit{true} spin operators $(\vec \Sigma,  \vec S)$ of both species. 

This expression can be developed into the various products of the spin operators components $\Sigma^{\pm,z} S^{\pm,z}$, defining $S^{\pm} = S^x \pm i S^y$ and $\Sigma^{\pm} = \Sigma^x \pm i \Sigma^y$. Two families of processes appear. \textbf{First,} relaxation and exchange terms that change the bath magnetization, $\Sigma^{\pm,z} S^{\pm}$. Here we consider a bath initially polarized by a magnetic field in the stretched state that minimizes the Zeeman energy. As illustrated in fig.\,\ref{fig:overview}, for a strong enough field  ($g_B \mu_B B \gg k_B T, J$), these processes will be either forbidden or negligible due to energy conservation. \textbf{Second,} relaxation and Ising terms that conserve the bath magnetization: $\Sigma^{\pm,z} S^z$. These terms are the ones driving the dissipative evolution of the spin chain. We will describe their effect using the Fermi golden rule, to evaluate the rate $\Gamma_{i \rightarrow f}$ of transfer between any two collective spin eigenstates $\ket{i_{spin}}, \ket{f_{spin}}$, associated with energy transfer to the bath density modes.
Writing $p_i$ ($p_f$) the probability that the spin chain occupies an eigenstate $\ket{i_{spin}}$ ($\ket{f_{spin}}$), all occupation probabilities will evolve following a set of equations
\begin{equation}
\frac{d p_i}{dt} = \sum_f (- \Gamma_{i \rightarrow f} \, p_i +\Gamma_{f \rightarrow i} \, p_f )
\end{equation}
where generally $\Gamma_{i \rightarrow f} \neq \Gamma_{f \rightarrow i}$. We now will evaluate the rate associated with each possible $i \rightarrow f$ process, based on the strength of the inter-species coupling $H_{int}$ and the bath density of states. 

\section{Analytical expression of rates}
\label{subsection:dipolargamma}

\subsection{Laying out the Fermi golden rule}
\begin{figure}
  \begin{center}
      \includegraphics[width=0.45\columnwidth]{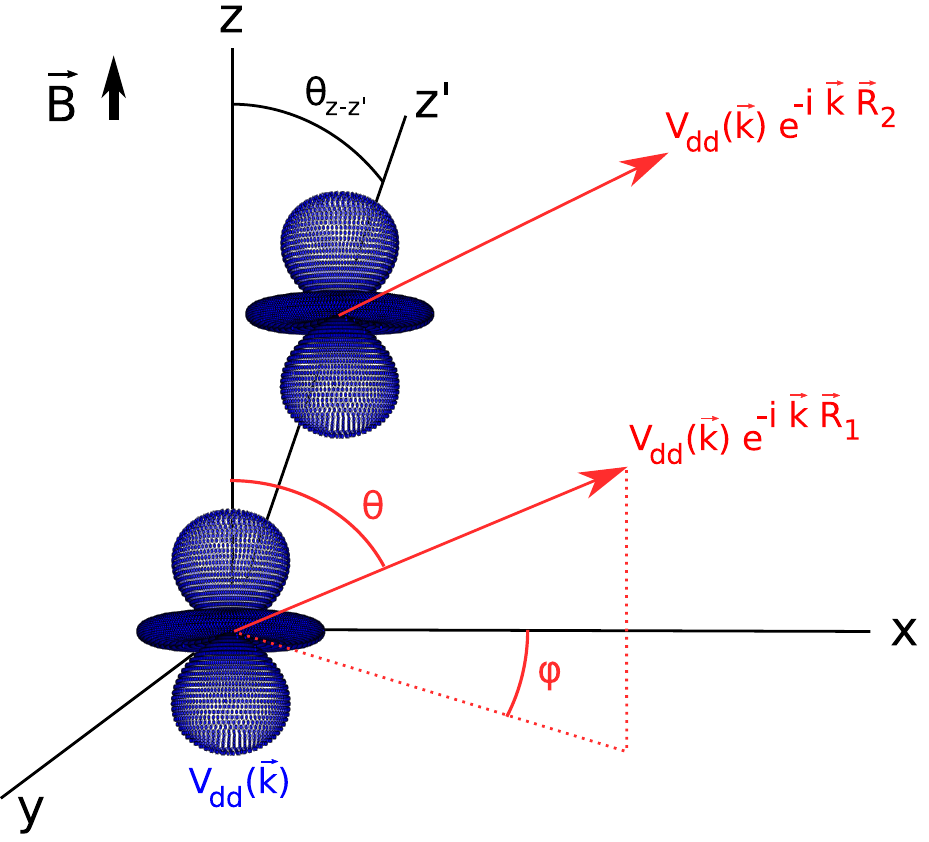}
      \caption{{\bf System geometry.} The quantization axis is defined by the
       magnetic field along $z$, and the spin chain is oriented along $z'$. The emission of one bath excitation with wavevector $\vec k$, with spherical angle coordinates $(\theta, \phi)$, acts on each spin in the chain with amplitude $V_{dd}(\vec k) \exp(- i \vec k \cdot \vec R_j)$. The angular dependence of $V_{dd}(\vec k)$ is illustrated as blue surfaces around each of the two spins (here for the $\Sigma^z S^z$ component of $V_{dd}$): the amplitude of $V_{dd}(\vec k)$ is given by the distance between surface and origin, along the direction of $\vec k$. Due to the excitation phases $ \exp(- i \vec k \cdot \vec R_j)$, emission  can change the chain spin correlations, except when $\vec k \perp \vec z\,'$. \label{fig:schema}}
  \end{center}
\end{figure}

The geometry of our system is as follows. The two atomic species are trapped by the same lattice laser beams, but see a different potential depth due to their different electronic structures. We will only consider situations where the fermions are effectively arranged into disconnected 1D chains, mostly out of computational time limits, while the bosons retain 3D coherence (see section \ref{subsec:bathlattice_params}). The formalism is nevertheless generic, and dissipation imposed on 2D or 3D lattice structures would deserve further scrutiny. The quantization axis $z$ is defined along the magnetic field. The spin chain axis $z'$ can differ from $z$ (see fig.\,\ref{fig:schema}).

Several restrictions are made for the sake of simplicity, that are not fundamental requirements. i) We consider that the external magnetic field is strong enough to maintain the bath polarization: $g_B \mu_B B > k_B T, J$. ii) We restrict ourselves to pseudo-spin 1/2 chains, where only two neighboring Zeeman states of the fermions are ever populated, and assume that the Zeeman energy difference between these two states $\Delta$ can be made negligible, $\Delta \ll J$, or irrelevant (see in section \ref{subsec:hierarchy} considerations on how to impose this). As $J \ll U$ this implies $\Delta\ll U$: all single-particle energies are smaller than $U$, such that with the proper chemical potential the occurrence of holes and doublons remains negligible.

We now describe the interaction between the dipolar bosons and the spin chain. In the fermionic Mott regime for large enough $U/t$, with one atom per site, we make the simplified assumption that $k_B T \ll U$ ensures negligible residual number fluctuations  \cite{DeLeo2008, Greif2016, Jordens2010, Cheuk2016b} (perfect Fock state on each site). % Mazurenko2017 Liu2006 Rigol2003, Boll2016
Under this approximation the density and spin operators for the fermions factorize, and we can label fermions by their site. The density distribution of the j-th fermion $n_F^j(\vec r)$ is constant, set by the ground band Wannier function on site $j$.
For the bosons, as the BEC is energetically constrained to the polarized state, spin and density also factorize. This enables us to write the coupling between spin chain and condensate as:
\begin{equation}
H_{int} = \sum_j \int d^3r d^3r' n^j_F(\vec r\,') \hat n_B(\vec r) V_{dd}^j(\vec r , \vec r\,')
\label{eq:veryfirstHint}
\end{equation}
Here, $\hat n_B$ is the bath density operator in second quantization. $V_{dd}^j$ is the interaction $V_{dd}$ of eq. \ref{eq:Vddr}, acting on the fermionic spin $\vec \Sigma_j$ of the fermion on the j-th site under the influence of a boson of spin $\vec S$ at position $\vec r$. It includes the added constraint of fixed BEC magnetization: the $S^+$ and $S^-$ terms are neglected, and the spin operator $S^z$ is replaced by the spin length $S$ of this species. We use Fourier transformation and write:
\begin{equation}
V_{dd}^j(\vec r , \vec r\,') = \int \frac{d^3k}{(2 \pi)^3} V_{dd}^j(\vec k) \exp\big(i \vec k \cdot (\vec r - \vec r\,')\big) 
\label{eq:veryfirstVdd}
\end{equation}
with
\begin{equation}
V_{dd}^j(\vec k) = \frac{\mu_0}{3} g_B g_F \mu_B^2 \left( 3 \frac{k_z}{k^2} (\vec \Sigma_j \cdot \vec k)- \Sigma_j^z \right) S  
\end{equation}
As in dipolar relaxation, total intrinsic spin is not conserved. The magnetization changing collisions are tied to the spin orbit coupling inherent to dipolar interactions \cite{Pasquiou2010}. 

We now describe the evolution of the spin chain under the assumption that the BEC behaves as a bath. Proper justification of this will be given once the formalism is entirely laid out, in section \ref{subsec:incoherentjustif}.
The processes that transfer energy from the spin chain to the BEC connect an initial spin state $\ket{i_{spin}} $ to a final spin state $\ket{f_{spin}}$ with energy difference $E_i-E_f = E_{if}$. Starting from an initial bath state $\ket{i_{bath}}$, the corresponding rates are:
\begin{eqnarray}
\Gamma_{i \rightarrow f} = \frac{2 \pi}{\hbar} \sum_{\ket{f_{bath}}} 
\lvert \bra{f_{spin}; f_{bath}} H_{int}  \ket{i_{spin}; i_{bath}} \rvert ^2 
~\delta(E_{if} + E_{if}^{bath})
\label{eq:fermirule}
\end{eqnarray}
This equation involves a summation over all possible final bath states $\ket{f_{bath}}$, with bath energy difference $E_{if}^{bath}$. We express all equations for a finite box of volume $V$ with periodic boundary conditions, so that this summation is discrete.
Regarding the bath, we start our derivation in the zero-temperature limit, only to generalize to finite T in the later section \ref{subsec:finiteT}. Quantum depletion is neglected (see section \ref{subsec:bathlattice_params}). Thus, the initial bath state is \textit{for now} described by a coherent state $\ket{BEC}$  with average atom number $N_0$.

\subsection{Matrix elements}
We now derive an explicit expression of the coupling element of eq. \ref{eq:fermirule}. Using eqs \ref{eq:veryfirstHint} and \ref{eq:veryfirstVdd}, we have:
\begin{eqnarray}
\bra{f_{spin}; f_{bath}} H_{int}  \ket{i_{spin}; i_{bath}} &=&
\sum_j \int \frac{d^3k_1}{(2 \pi)^3} \bra{f_{spin}} V_{dd}^j(\vec k_1)\ket{i_{spin}}\nonumber \\
~&~& \cdot \int_V d^3r' e^{-i \vec k_1 \cdot \vec r\,'} n_F^j(\vec r\,') 
\nonumber \\ ~&~&
\cdot \int_V d^3r \bra{f_{bath}}e^{i \vec k_1 \cdot \vec r} \hat n_B(\vec r) \ket{BEC}
\end{eqnarray}
There, the corresponding physical process is a transfer of momentum $\vec k_1$ from the fermions to the bosons. The lattice potential % and the Mott regime
constrains the fermions to their unchanged Wannier states, whose finite momentum widths allow for this momentum transfer : $n_F^j(\vec r\,') = \mid {\rm w_F}(\vec r\,' - \vec R_j) \mid^2$, where ${\rm w_F} (\vec r\,' - \vec R_j)$ is the ground band Wannier function centered on the site coordinate $\vec R_j$.
We calculate the bath matrix element using the discrete basis of plane waves in cubic volume $V$ with periodic boundary conditions: $\hat n_B(\vec r) =   \sum_{\vec k_2} \sum_{\vec k_3} \frac{e^{i (\vec k_3 - \vec k_2) \cdot \vec r}}{V} \psi^\dagger(\vec k_2)\psi(\vec k_3)$. 
Writing ${\rm F}[f](\vec k)$ the Fourier transform of a function $f(\vec r)$, we obtain:
\begin{eqnarray}
\bra{f_{spin}; f_{bath}} H_{int}  \ket{i_{spin}; i_{bath}} =
\int \frac{d^3k_1}{(2 \pi)^3} \sum_j e^{-i \vec k_1 \cdot \vec R_j} \bra{f_{spin}}  V_{dd}^j(\vec k_1)\ket{i_{spin}}   
\nonumber \\ 
\cdot ~ {\rm{F[|w_F|}^2]} (\vec k_1)  
\sum_{\vec k_2} \bra{f_{bath}} \psi^\dagger (\vec k_2) \psi(\vec k_2 - \vec k_1) \ket{BEC}  
\label{eq:developcoupling1}
\end{eqnarray}
This expression explicitly relates the change in collective spin state to the interference of phonon radiation patterns from the different fermions, as an operator acting {\it collectively} on all the spins of the chain appears: 
\begin{equation}
V_{col}(\vec k_1) = \sum_j e^{-i \vec k_1 \cdot \vec R_j} V_{dd}^j (\vec k_1)
\end{equation}
(see fig.\,\ref{fig:schema}). The operator $V_{col}$ defines selection rules regarding which collective spin states are coupled.
For these rules, we diagonalize the effective Hamiltonian $H_{mag}$ together with the total pseudo-spin length operator $\vec \Sigma_{eff, tot}^2 = (\sum_j \vec \Sigma_{eff,j})^2 $ and with both the total spin and pseudo-spin projection operators, e.g. $\Sigma_{tot}^z = \sum_j \Sigma^z_j$. In this eigenstate basis, if $\ket{i_{spin}}$ and $\ket{f_{spin}}$ differ by up to one in pseudo-spin length $\Sigma_{eff,tot}$ and 0 in projection $\Sigma_{tot}^z$,
\begin{eqnarray}
\bra{f_{spin}}  V_{col}(\vec k)\ket{i_{spin}}_{\Delta \Sigma_{tot}^z = 0} \propto (3 \frac{k_z^2}{k^2} - 1) 
\cdot \sum_j  e^{-i \vec k \cdot \vec R_j} \bra{f_{spin}} \Sigma^z_j\ket{i_{spin}} 
\label{eq:piVcol}
\end{eqnarray}
and the corresponding two-body operator is $\Sigma^z S^z$. If $\ket{i_{spin}}$ and $\ket{f_{spin}}$ differ by up to one in pseudo-spin length $\Sigma_{eff,tot}$ and $\pm 1$ in projection $\Sigma_{tot}^z$,
\begin{equation}
\bra{f_{spin}}  V_{col}(\vec k)\ket{i_{spin}}_{\Delta \Sigma_{tot}^z = \pm 1}\propto  3 \frac{k_z}{k^2} \frac{k_x \mp i k_y}{2}   
 \cdot \sum_j e^{-i \vec k \cdot \vec R_j} \bra{f_{spin}} (\Sigma^x_j \pm i \Sigma^y_j) \ket{i_{spin}} 
\label{eq:sigVcol}
\end{equation}
and the corresponding  two-body operator is $\Sigma^\pm S^z$.
Otherwise, the coupling vanishes.

\subsection{Bath description}
\label{subsec:bathdescription}
The coupling term (eq.\,\ref{eq:developcoupling1}) is sensitive to the potentials seen by the bosons. 
Here, we assume that the bosons in the lattice retain large coherence over the 3D sample, which we check independently (see section \ref{subsec:bathlattice_params}). 
We develop accordingly the BEC matrix element in equation \ref{eq:developcoupling1} by transforming annihilation and creation operators from the plane wave basis $\psi(\vec q)$ to the Bloch wave basis $\phi(\vec q)$, and then to the Bogoliubov excitation basis $b(\vec q)$ and BEC state $\hat c_{BEC}$.

The first transformation, between plane waves and Bloch waves, is based on the decomposition of the Bloch band structure of our lattice, of axes $(x', y', z')$. We assume a cubic lattice with same period along all three axes, and reciprocal lattice wavevector length $K$. Then,
\begin{eqnarray}
\phi(\vec q) = \sum_{n_x \in \mathbb{Z}} \sum_{n_y \in \mathbb{Z}} \sum_{n_z \in \mathbb{Z}} a_{n_x}^*[q_{x'}] a_{n_y}^*[q_{y'}] a_{n_z}^*[q_{z'}] ~ \psi(\vec q + \vec n K)
\label{eq:textbookBloch}
\end{eqnarray}
where $\vec q = q_{x'} \vec e_{x'} + q_{y'} \vec e_{y'} +  q_{z'} \vec e_{z'}$ is in the first Brillouin zone, and we define the notation $\vec n = n_x \vec e_{x'} +n_y \vec e_{y'} +n_z \vec e_{z'}$. The lattice can have different depths along the three axes, such that the $a_{n_x}, a_{n_y}, a_{n_z}$ implicitly differ. We omit in our notations the band indices: in this work only the ground band is ever populated given the relevant energies. We generalize the notation: $a_{\vec n}(\vec q) = a_{n_x}[q_{x'}] a_{n_y}[q_{y'}] a_{n_z}[q_{z'}]$.  

The second step is the Bogoliubov transformation \cite{Pitaevskii2016}: 
\begin{equation}
\phi(\vec q) = u(\vec q) b(\vec q) + v^*(- \vec q) b^\dagger (-\vec q) + \delta^K(\vec q, \vec 0) \hat c_{BEC}
\label{eq:textbookBogo}
\end{equation}
where $\delta^K$ is a Kronecker symbol. The prefactors $(u(\vec q),v^*(-\vec q))$ depend on kinetic and interaction terms ($E_{\vec q}, V_{\vec q}$ respectively) that can be strongly affected by the lattice potential:
\begin{eqnarray}
u(\vec q), v(\vec q) = \pm \sqrt{\frac{1}{2} \left(\pm1 + \frac{E_{\vec q} + V_{\vec q}}{\sqrt{E_{\vec q} (E_{\vec q} + 2 V_{\vec q})}}\right)} 
\end{eqnarray}
$E_{\vec q}$ is the non-interacting energy for wavevector $\vec q$ of the Bloch wave. The interaction term $V_{\vec q}$ depends on wavevector due to the dipolar interaction within the polarized dipolar condensate. 
In the absence of a lattice potentials for bosons, $V_{\vec q} = g \rho_0 + \frac{\mu_0}{3}(g_B \mu_B S)^2 f(\vec q) \rho_0$, with $\rho_0 = N_0/V$ the mean bosonic density, $g$ the 3D pseudopotential interaction parameter, and $f(\vec q) =  (3 q_z^2/q^2 -1)$. 
We generalize this in the presence of the lattice potential seen by the bosons, using a tight binding approximation: we introduce boson Wannier functions ${\rm w_B}^j(\vec r) = {\rm w_B}(\vec r - \vec R_j)$  on each site $j$, and assume that ${\rm w_B}^{j,*}(\vec r) {\rm w_B}^{k}(\vec r)$ can be approximated by $\delta^K(j,k) \lvert {\rm w_B}^j(\vec r) \rvert^2$.
We obtain: 
\begin{equation}
V_{\vec q} = \tilde g \rho_0 + \frac{\mu_0}{3}(g_B \mu_B S)^2 \rho_0  \sum_{\vec n\in \mathbb{Z}^3} \bigg \lvert {\rm{F[|w_B|}^2]} (\vec q + \vec n K) \bigg \rvert ^2 f(\vec q + \vec n K)
\label{eq:tightbindingVq}
\end{equation} 
where $\tilde g$ is the bath contact interaction parameter renormalized by the lattice. For a 3D lattice confinement such that the bosons experience on-site interaction $U_B$, this interaction parameter reads $\tilde g =  U_B (\lambda/2)^3$, with $\lambda/2$ the lattice spacing. The Bogoliubov quasi-particles have an anisotropic dispersion relation $\epsilon(\vec q)= \sqrt{E_{\vec q} (E_{\vec q} + 2 V_{\vec q})}$.

We finally evaluate the coupling element eq. \ref{eq:developcoupling1}, using the basis transformations of bosonic operators, equations \ref{eq:textbookBloch} and \ref{eq:textbookBogo}. 
The coupling Hamiltonian $H_{int}$ is of second order in boson field operators, which we just decomposed between Bogoliubov modes and BEC state mode. Consequently, the initial and final bath states can only be coupled if they differ by zero, one or two Bogoliubov annihilation and creation operators. The dominant terms for dissipation are those differing by one excitation, which also contain one BEC operator with macroscopic expectation value $\sqrt{N_0}$. Thus, for $T=0$ and neglecting quantum depletion, the state $\ket{f_{bath}}$ only differs from $\ket{BEC}$ by the creation of a Bogoliubov excitation, parametrized by its wavevector $\vec q$:
\begin{equation}
\ket{f_{bath}(\vec q)} = b^\dagger(\vec q) \ket{BEC}
\end{equation} 
We assume the BEC to occupy the zero-momentum Bloch wave state. Then, we get \textit{without} explicit tight binding approximation
\begin{eqnarray}
\bra{f_{spin}; f_{bath}(\vec q)} H_{int}  \ket{i_{spin}; i_{bath}} = 
 \frac{\sqrt{N_0}}{V} \sum_{\vec n\in \mathbb{Z}^3} \sum_{\vec m\in \mathbb{Z}^3} 
 \bra{f_{spin}} V_{col}( \vec q + \vec m K)\ket{i_{spin}}  
 \nonumber \\ 
\cdot \Big( 
a^*_{\vec n}[\vec q] a_{\vec n- \vec m}[0] u^*(\vec q)
 + a^*_{\vec n}[0] a_{\vec n- \vec m}[-\vec q] v^*(\vec q)
\Big) ~ {\rm{F[|w_F|^2]}} ( \vec q + \vec m K ) 
\label{eq:developcoupling2}
\end{eqnarray}
The tight binding approximation may also be applied along some dimensions to simplify eq.\,\ref{eq:developcoupling2}. Applying it to all three dimensions leads to
\begin{eqnarray}
\bra{f_{spin}; f_{bath}(\vec q)} H_{int}  \ket{i_{spin}; i_{bath}} = 
 \frac{\sqrt{N_0}}{V}  (  u^*(\vec q) + v^*(\vec q)) \nonumber \\
 \cdot  \sum_{\vec n\in \mathbb{Z}^3} \bra{f_{spin}} V_{col}( q + \vec n K)\ket{i_{spin}}  
 {\rm{F[|w_F|^2]}} ( q + \vec n K )\cdot  {\rm{F[|w_B|^2]}} ( - \vec q -\vec n K  )
\label{eq:developcoupling2_tightbinding}
\end{eqnarray}
From equation \ref{eq:developcoupling2_tightbinding} we can see that the concentration of bosons onto Wannier functions by the lattice has an impact on the inter-species dipole coupling \footnote{For example, if the boson lattice sites are located half a fringe away from the fermion lattice sites due to opposite potentials, the coupling is generally reduced. Indeed, the expression \ref{eq:developcoupling2_tightbinding} acquires a phase inside the sum, $e^{i (\vec q + \vec n K)\frac{\pi}{K} (\vec e_{x'} + \vec e_{y'} + \vec e_{z'}) }$ of sign alternating with $\vec n$.}, despite the long-range nature of the dipolar interaction.
Due to the anisotropic nature of $V_{col}$, the lattice can be beneficial or harmful to the coupling strength, as the reciprocal space summation $\sum_{\vec n}$ can involve terms of opposite signs, depending on the weights imposed by the Wannier functions.

\subsection{Zero-temperature rates}

At this stage we conclude our calculation for $T=0$. To reveal explicitly the role of box volume, we replace the discrete summation on $\ket{f_{bath}(\vec q)}$ by a continuous one on $\vec q$.
\begin{eqnarray}
\Gamma_{i \rightarrow f}^{T=0} &=& \frac{2 \pi}{\hbar}
 \int_{\mathrm{Brillouin\,Zone}} d^3q \frac{V}{(2 \pi)^3} \delta(\epsilon(\vec q) - E_{if}) 
\lvert  \bra{f_{spin}; f_{bath}(\vec q)} H_{int}  \ket{i_{spin}; i_{bath}}   \rvert ^2 
\nonumber \\
 ~&=& \frac{2 \pi}{\hbar}
 \int_0^\pi sin(\theta)d\theta \int_0^{2 \pi} d\phi\, q^2(\theta, \phi, E_{if}) \frac{V}{(2 \pi)^3}\frac{1}{\vert \partial \epsilon(\vec q) / \partial q ~_{\vert \theta, \phi} \vert}\nonumber \\
 ~&~&\cdot \lvert  \bra{f_{spin}; f_{bath}(\vec q)} H_{int}  \ket{i_{spin}; i_{bath}} \rvert ^2 
\label{eq:dipolarfermirule}
\end{eqnarray}
where the 3D Dirac function results in an integration on the 2D anisotropic wavevector surface $\vec q(\theta, \phi, E_{if})$ defined by conservation of energy: $\epsilon(\vec q) = E_{if}$. In  this paper, these energies will always lie in the ground band for the bosons.

\subsection{Finite temperature rates}
\label{subsec:finiteT}
At finite temperature, the initial bath state is not a pure state anymore. The theory above can then be generalised. We use a bath eigenstate basis $\ket{\{n_q\}} = \prod_q (b^\dagger_q)^{n_q} \ket{BEC}$ consisting of Fock states of the Bogoliubov modes labelled $q$ and of energy $\epsilon_q$. 
The statistical probability of any such eigenstate $\ket{\{n_q\}^{\rm init}}$ is 
\begin{equation}
{\rm p}(\{n_q\}^{\rm init}) = \frac{e^{- \sum_q \epsilon_q n_q^{\rm init} / k_B T}}
{\sum_{\{n_q \}} e^{- \sum_q \epsilon_q n_q / k_B T}}
\end{equation}
The rate between two spin chain states $\ket{i_{spin}}, \ket{f_{spin}}$ with respective energies differing by $E_i - E_f = E_{if}$ now writes:
\begin{equation}
\Gamma_{i \rightarrow f} = \sum_{\{n_q \}^{\rm init}} {\rm p}(\{n_q\}^{\rm init})
\sum_{\{n_q \}^{\rm final}} \delta(E_{if} + E_{if}^{bath}) %\nonumber \\
 \Big\lvert \langle \{n_q \}^{\rm final} ; f_{spin} \vert H_{int} \vert \{n_q \}^{\rm init} ; i_{spin} \rangle \Big\rvert ^2
\label{eq:finiteTgoldenrule}
\end{equation}
where $E_{if}^{bath} = \sum_q (n_q^{\rm init} - n_q^{\rm final})  \epsilon_q$. As explained in section \ref{subsec:bathdescription}, the dominant terms in the decomposition of $H_{int}$ are those involving one Bogoliubov operator and the BEC mode.

\textbf{If the spin energy diminishes ($E_f < E_i$)}, the most contributing $\ket{\{n_q\}^{\rm final}}$ differ from $\ket{\{n_q\}^{\rm init}}$ by one added excitation (phonon emission). Bose enhancement factors appear as excitation states are not necessarily initially empty: $n_q^{\rm init}\geq 0$. After statistical averaging over $\ket{\{n_q\}^{\rm init}}$, as explicit in eq. \ref{eq:finiteTgoldenrule}, the rate $\Gamma_{i \rightarrow f}$ is modified by finite temperature $T$ through an additional factor:
\begin{equation}
\Gamma_{i \rightarrow f}^{T > 0} = \Gamma_{i \rightarrow f}^{T = 0} \cdot \langle n_q + 1 \rangle
\label{eq:cool}
\end{equation}
where $\langle n_q \rangle = \frac{1}{e^{E_{if}/k_B T} -1} $  depends on the spin energy released in the phonon.

\textbf{If the spin energy increases ($E_f > E_i$)}, the most contributing $\ket{\{n_q\}^{\rm final}}$ differ from $\ket{\{n_q\}^{\rm init}}$ by one less excitation. The rate $\Gamma_{i \rightarrow f}$, associated to phonon absorption, is now non-zero. It scales with the inverse (phonon emission) process as:
\begin{equation}
\Gamma_{i \rightarrow f}^{T > 0} = \Gamma_{f \rightarrow i}^{T = 0} \cdot \langle n_q \rangle
\label{eq:heat}
\end{equation}

This assymetry between the processes pumping energy out (eq. \ref{eq:cool}) or in (eq. \ref{eq:heat}) the spin chain leads to thermalization between the spin chain collective degrees of freedom and the bath motional temperature. This is of particular interest for a sufficiently cold BEC, $k_B T \lesssim J$, such that $\langle n_q \rangle \lesssim 1$.

\subsection{Validity of the bath approximation and Fermi Golden rule}
\label{subsec:incoherentjustif}
We conclude this section by defining properly the initial assumption of treating the BEC as a bath, (i) with which energy exchange processes are irreversible (justifying the Fermi golden rule), and (ii) such that its state is negligibly affected by interaction with the chain. 

(i) For a given process $i \rightarrow f$, dephasing of the populated phonon modes is required to prevent coherent oscillations of the energy between chain and bath. As for spontaneous emission of light, this is satisfied even for an isolated BEC provided the number $N_{modes}$ of excitation modes with energies close enough from  $E_{if}$ is much larger than one. We estimate $N_{modes}$ by integrating the shell of wavevectors corresponding to excitations resonant with $E_{if}$, within an energy bandwidth given by the coupling strength:
\begin{equation}
N_{modes} \simeq \frac{2 \pi}{\hbar}
 \int_{\mathrm{Brillouin\,Zone}} d^3q \frac{V}{(2 \pi)^3} \delta(\epsilon(\vec q) - E_{if}) 
\lvert  \bra{f_{spin}; f_{bath}(\vec q)} H_{int}  \ket{i_{spin}; i_{bath}}   \rvert  
\end{equation}
This amounts to a product of density of state and coupling element, taking into account the anisotropy of the coupling.
For fixed BEC mean density $N_0/V$, matrix elements of $H_{int}$ scale as $1/\sqrt{V}$, such that $N_{modes}$ scales as $\sqrt{V}$ and incoherent dynamics is always justified at the thermodynamic limit. Once in the incoherent regime, the rate is independent on volume.
For small clouds that would not satisfy this density-of-state criterion, irreversibility requires dissipation or damping of the phonons by other means, either inherent to finite temperature BECs \cite{Jin1997, Giorgini1998, Williams2001, Mendonca2018}, or external such as evaporation. 
The latter is experimentally relevant, as BEC temperatures lower than $J$ are typically obtained in traps with low depths.

(ii) Furthermore, throughout the numerical results, the bath excitations will be assumed to be thermally distributed at a fixed temperature. Precisely, we assume that the emissions and absorptions of phonons affect negligibly the populations $\langle n_q \rangle$. For a strongly damped bath, this may be ensured by a sufficient heat capacity. For weakly damped baths as most BECs, this will be ensured either by evaporation, or by $N_{modes}$ being larger than the number of phonons emitted within the same energy band. The latter criterion improves with the ratio of boson atom number to fermion atom number.

%%%%%%%%%%%%%%%%%%%%%%%%%%%%%%%

\section{Thermalization of a spin chain}
\label{subsec:numerical}

\subsection{Spin chains in K-Dy mixtures}

Using one of the most magnetic dipolar species (Cr, Er, Dy) as the bath, dipolar interactions will drive dissipation even on spin chains from the weakly dipolar alkali species. More exotic situations with actual dipole-dipole interactions \textit{within} the spin chain could be studied, e.g. using an isotope of the strongly dipolar species for the chain as well. 

In the prospect of studies of the t-J model, constituting spin chains from \textit{fermionic} species in the Mott regime is very attractive. Lithium or Potassium are the most prominent in recent experiments, together with alkaline-earth species that however have much too weak Land\'e factors in the ground state for our present purpose. \textbf{We will here discuss $^{40}$K in its ground hyperfine state}, with spin $F = 9/2$, and Land\'e factor $g_F = 2/9$. Fermionic Lithium would be more problematic due to a combination of complications \footnote{On the one hand, at low magnetic field the weak interactions of $^6$Li are unsuited for the emulation of magnetism. On the other hand, Lithium Feshbach resonances are located in the Paschen-Back regime. There, the lowest-lying atomic states differ mostly by the \textit{nuclear} spin state, with too small Lande factor. For state combinations differing by their electronic spin state, relaxation of the large Zeeman energy would need to be prevented 
\cite{Pasquiou2010, Pasquiou2011a}.}.

\begin{figure}
  \begin{center}
      \includegraphics[width=0.9\columnwidth]{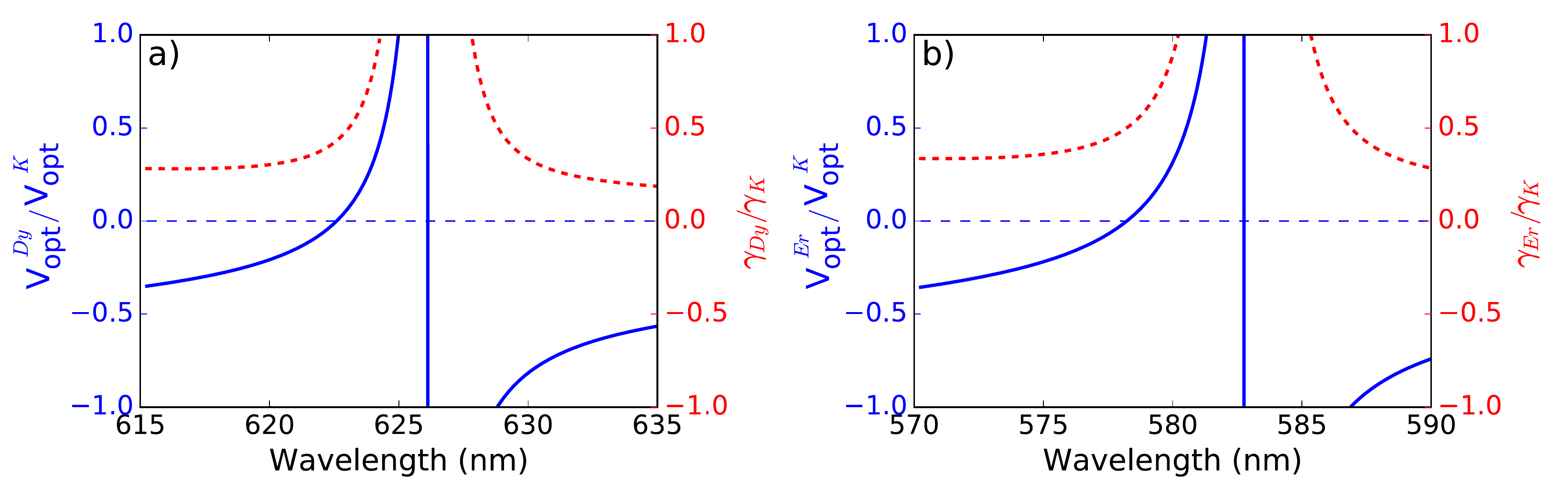}
      \caption{{\bf Lattice wavelength selection}. The ratio of the optical potentials $V_{opt}$ (continuous lines) and spontaneous light emission rate $\gamma$ (dashed lines) associated to both bath and spin chain species are shown for K-Dy mixtures (a) and K-Er mixtures (b), in the vicinity of narrow lines of the bath species in the visible range. Those lines offer within a few nanometers a large tunability of the bosonic lattice depth with sufficiently small spontaneous emission.
      \label{fig:latticelambda}}
  \end{center}
\end{figure}

A key point is the optical lattice potential seen by the bath species. It generally differs from that seen by the spin chain species. As mentioned earlier (section \ref{subsec:bathdescription}), a lattice of opposite sign may often lead to poor interspecies dipolar coupling, concentrating the bosons half-a-space away from the fermions. 
For Dy or Er in association with K spins, wavelengths $> 767$ nm would generate lattices of the same sign for both species. However larger $J$ \textbf{and} large interspecies coupling can be obtained at shorter visible wavelengths, using the red-detuned vicinity of "narrow" lines that would dominate over the effect of the wide lines at 421 and 401 nm respectively for Dy and Er. We suggest for example one narrow line of Dy at 626.1 nm (140 kHz) and one narrow line of Er at 582.8 nm (190 kHz) (see figure \ref{fig:latticelambda}).
The intervals 622.5 - 624.3 nm for Dy and 578.2 - 580.2 nm for Er are attractive, as in these intervals the boson lattice depths can be tuned from 0 to $\sim$0.4 times the fermion lattice depth while light scattering remains dominated by the fermions. Here we will use $^{164}$Dy, which has the strongest magnetic dipole moment $\mu = g_B S \mu_B  = 10 \mu_B$, but the shorter lattice wavelengths suggested for Er translates in higher super-exchange and fairly similar spin cooling rates despite the lower moment $7 \mu_B$. 

In the following, we will consider $^{40}$K in the presence of $^{164}$Dy\,\cite{Ravensbergen2018}, in an optical lattice with wavelength $\lambda_L \approx 623\,$nm. We effectively create a 2D array of 1D spin chains by using a lattice, as seen by K, of depth $25\,E_r^K$ along two axes $x',y'$ (where $E_r^K = h^2 /(2 m_K \lambda_L^2)$ is the potassium recoil energy), and $3.5\,E_r^K$ along one axis $z'$. The spontaneous light emission rate is $\sim$ 0.2\,/s. 
The on-site interaction energy $U = 7.5 \, t_{z'}$, where $t_{z'}$ is the tunnelling energy along the weak axis $z'$. Fermions can then be prepared in the 1D Mott regime with unit filling \cite{Rigol2003, Rigol2004, Boll2016}. 
The super-exchange energy is, along the weak axis, $J = h \times 632\,$Hz, while it is along the transverse axes $h \times\,0.07\,$Hz, thus effectively creating decoupled 1D spin chains. The lattice depth for the bosons, on the other hand, sensitively depends on the lattice wavelength within a few nanometers, and will be used as an independent optimization variable in section \ref{subsec:bathlattice_params}. 

\subsection{Hierarchy of energy scales}
\label{subsec:hierarchy}
The actual implementation requires to respect a certain hierarchy of energy scales, involving the magnetic field $B$, the BEC temperature $T$, and the superexchange energy of fermions $J$ (see fig.\,\ref{fig:overview}).
\textbf{First,} the bath temperature $T < J/k_B \simeq 30\,$nK  \footnote{This is within the bounds of state-of-the-art BECs in optical lattices. Already in bulk, 1\,nK is achieved in experiments \cite{Olf2015}. The lowest temperatures are achieved using shallow traps and eventually adiabatic decompression, which is consistent with the low BEC density specified in the later section \ref{subsec:bathlattice_params}. Furthermore, loading optical lattices can significantly reduce BEC temperatures \cite{Rey2006}. The work on thermometry of BECs in lattices \cite{Trotzky2010} shows temperatures down to 3\,nK.} is such that low energy spin chain states are favored by coupling to the bath. 
\textbf{Second,} $(J, k_B T) < g_B \mu_B B$ ensures that the bath remains polarized. This is largely satisfied with fields in the mG range. Although the scheme might work also for a BEC with free magnetization, this study is beyond the scope of our calculations. \textbf{Third,} to study antiferromagnetic models the fermion Zeeman shift $\Delta$ should not favor trivially polarized spin chains. There are several ways to ensure this, with different consequences.

i) The fermionic Zeeman states of interest may be kept quasi-degenerate by $ g_F \mu_B B \ll J$. Combined with our prior assumption of a polarized BEC, this implies $g_F \ll g_B$, which is not well satisfied for most of the atomic species combinations. To circumvent this, we propose to add microwave dressing to the spin states \cite{Gerbier2006}: the linear Zeeman effect can be combined with an engineered quadratic shift $\alpha m_F^2$, such that the detuning between the two lowest Zeeman states is $\Delta = \big\vert E_{m_F = -F} - E_{m_F = -F+1} \big\vert = \big\vert g_F \mu_B B - \alpha (F^2 - (F-1)^2) \big\vert \ll J$. To emulate pseudo-spin 1/2 dynamics in a larger spin species, one further requires $\big\vert E_{m_F = -F+1} - E_{m_F = -F+2} \big\vert > J, k_B T$ to prevent population of the other Zeeman states. This implies  $B >  \frac{J}{g_F \mu_B} \frac{2 F -1}{2} \simeq 8\,$mG.  This protocol, and the effect of a residual $\Delta$, are discussed further in Appendix A.  The calculations below assume such a setting. Varying $\Delta$ would enable studies of the Heisenberg anti-ferromagnetic model with free magnetization for varying bias field, complementing the more usual fixed magnetization setting in experiments.

ii) Another possibility is to ensure that the energy released by magnetization-changing collisions, $\Delta$, lies in the energy gaps within the boson band structure. Dipolar relaxation is then  prevented by energy conservation, as seen in \cite{Pasquiou2010, Pasquiou2011a}. 
The thermalization of the spin chain would then occur \textit{at fixed magnetization} (i.e. fixed spin projection along $\vec B$).
This approach requires a much lower degree of field control, and in particular releases the constraint that $\Delta \ll J$. Cooling of the chain is slightly less efficient, as the magnetization-changing processes are suppressed. Nevertheless, processes given by eq.\,\ref{eq:piVcol} are still at play and may thermalize the collective spin at fixed magnetization. The magnetization is set initially by creating the appropriate spin imbalance using standard manipulation techniques.

\subsection{Optimizing the bath lattice potential}
\label{subsec:bathlattice_params}
\begin{figure}
  \begin{center}
      \includegraphics[width=1 \columnwidth]{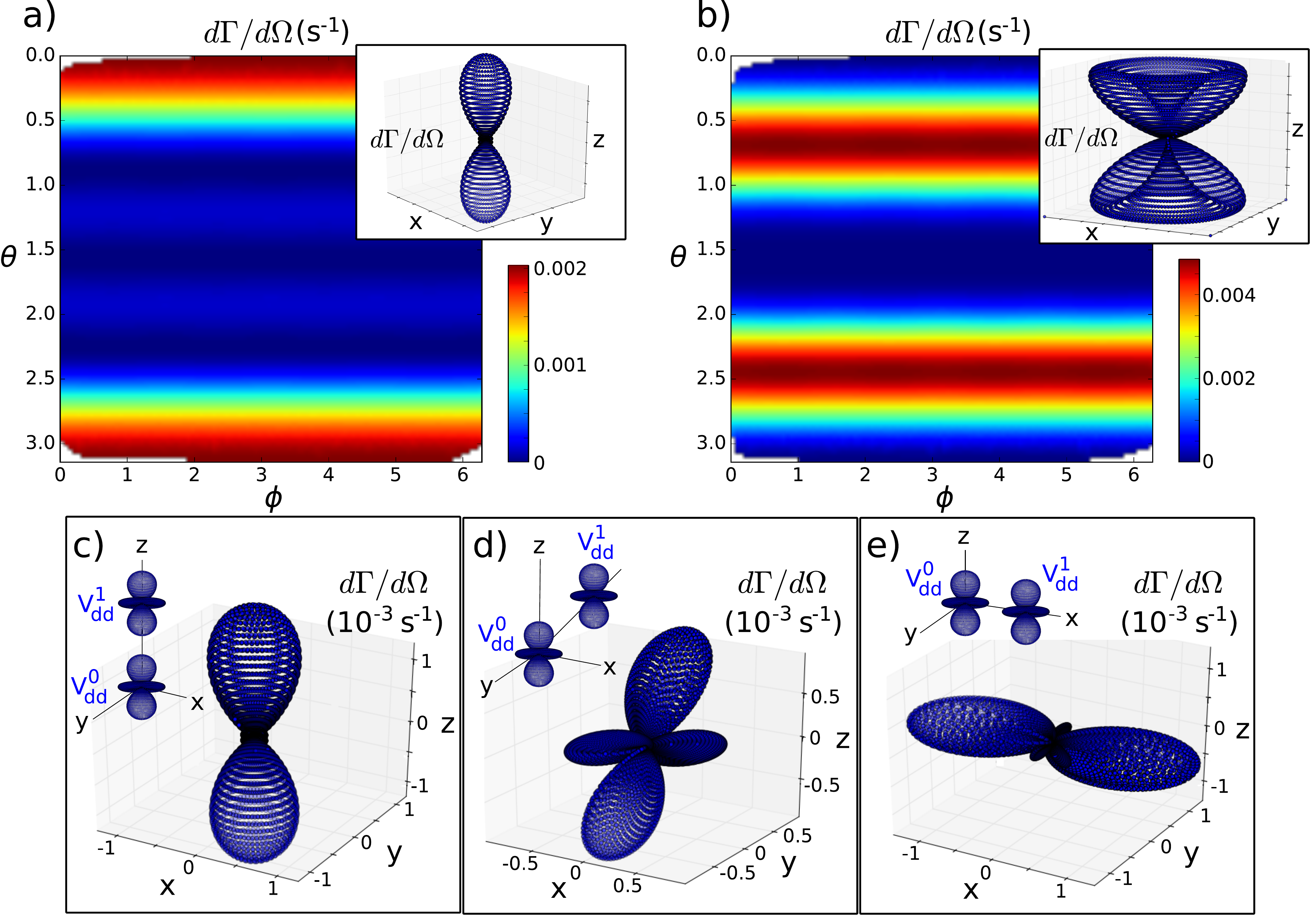}
      \caption{{\bf Phonon emission diagram} (rate per solid angle) as a function of phonon wavevector orientation $(\phi, \theta)$ for a length-two spin chain in the triplet states  \textbf{in a bath seeing no lattice}. The bath is polarized along the vertical $z$ axis. 
     (a): from $\ket{T_{0}}$, for $\theta_{z-z'} = 0$. The inset is a 3D representation of the emission diagram. 
     (b): from $\ket{T_{\pm 1}}$, for $\theta_{z-z'} = 0$.
     (c-d-e):  from $\ket{T_{0}}$, varying  $\theta_{z-z'} = 0, \pi/4, \pi/2$.  
Parameters: K-Dy mixture, 623 nm lattice with depth for fermions $(25, 25, 3.5) E_r^K$ along $(x',y',z')$, bath mean density $\rho_0 = 3. 10^{19}$/m$^3$, released super-exchange energy $J = h \cdot 632\,$Hz, and bath temperature $k_B T = 0.3 J$.
\label{fig:radiationdiagram}}
  \end{center}
\end{figure}

The lattice potential experienced by the bosons has a crucial effect on phonon emission, strongly tied to the anisotropic character of the dipolar interaction. We first illustrate the importance of this anisotropy without the lattice potential. Figure \ref{fig:radiationdiagram} presents phonon emission diagrams for different orientations of a 2-fermions chain in the triplet states. We make two observations on the emission directionality. (a) It is strongly spin dependent, as can be inferred by equations \ref{eq:piVcol} and \ref{eq:sigVcol}. (b) It is suppressed orthogonally to the spin chain, as the phase terms $e^{-i \vec q \cdot \vec R_j}$ in $V_{col}$ are close to 1. Indeed, $V_{col}$ then produces an identical rotation of all spins in the chain, which does not modify the collective spin and therefore is unable to drive any dissipative dynamics. 
\footnote{The interaction Hamiltonian may also drive collective precession of the chain under the influence of the magnetic field radiated by the bath dipoles, $\vec B_{int}$.  
For a finite size condensate with spherical symmetry, $\vec B_{int}=0$ at the center but is not spatially homogeneous. Orders of magnitude for the Larmor frequency of off-centered chains may lie in the 10\,Hz range \cite{Pasquiou2011b, Swislocki2011}. The chain eigenstates with same total spin length but different $z$ projection are coupled and degeneracies are lifted. Considering chains much smaller than the condensate, our expression for rates $\Gamma_{i \rightarrow j}$ in this paper remains valid in the new (local) eigenstate basis. The numerical results shown are for a centered chain, with no degeneracy lifting.
}

\begin{figure}
  \begin{center}
      \includegraphics[width=\columnwidth]{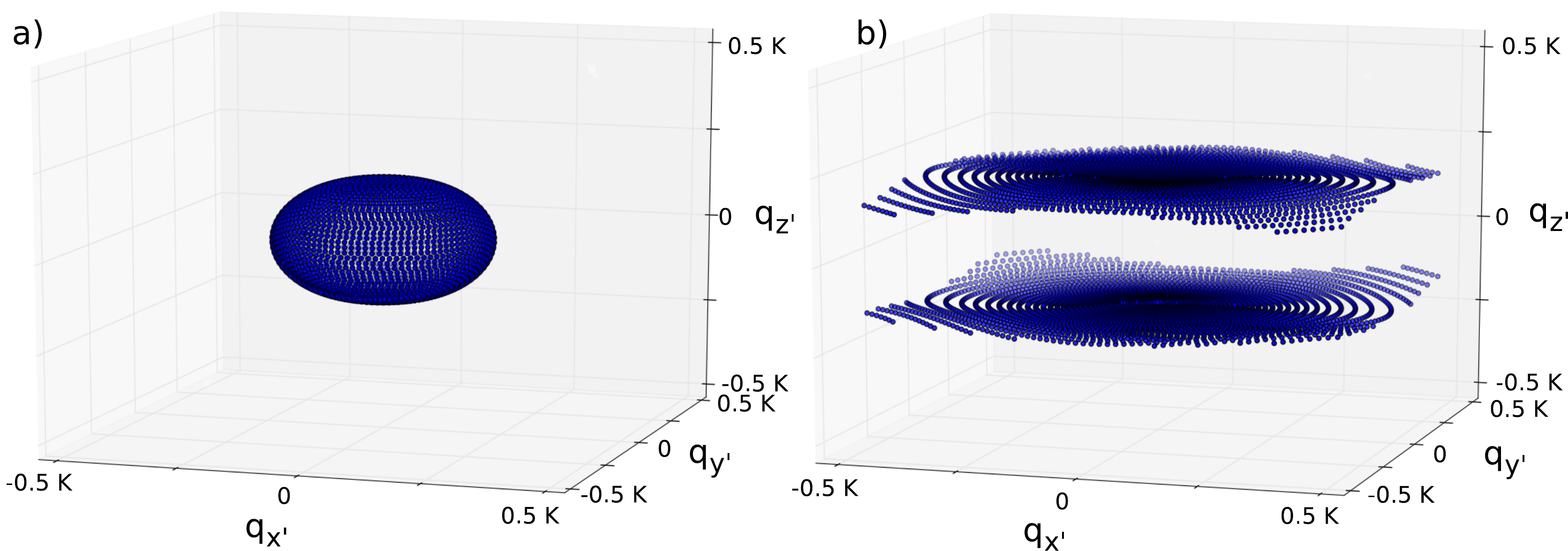}
      \caption{{\bf Resonant phonon wavevectors $\vec q(\theta, \phi, J)$, (a) without and (b) with  anisotropic bath lattice potential.}   Parameters as in fig.\,\ref{fig:radiationdiagram}, except for (b) a bath lattice potential, set to depths $(12, 12, 3.5) E_r^{Dy}$ along $(x',y',z')$. The sampling of wavevectors is detailed in Appendix B.
      \label{fig:dispersion}}
  \end{center}
\end{figure}

The bath lattice potential introduces the following effects. (i) It can strongly influence the anisotropy of the phonon dispersion relations, and thus of the resonant phonon wavevector surface (see figure \ref{fig:dispersion}). This affects both the density of states and the phase terms $e^{-i \vec q \cdot \vec R_j}$ in $V_{col}$ discussed above. (ii) It can generate either constructive or destructive interference through $\sum_{\vec n} [..] V_{col} (\vec q + \vec n K)$ terms of eqs. \ref{eq:developcoupling2} and \ref{eq:developcoupling2_tightbinding}, as the dipolar interaction changes sign depending on the orientation of $\vec q + \vec n K$. 

The lattice potential depth for bosons is thus an important control tool, with two main functions: i) tuning the boson dispersion relation, to explore high-density-of-states, and also to stabilize against the bath dynamical instabilities of strongly magnetic species \cite{Giovanazzi2006, Muller2011} (for Dy, $\epsilon_{dd} = \mu_0 \mu^2 m / 12 \pi \hbar^2 a_{sc} \simeq 1.4$); ii) maximizing the constructive interference and thus enhancing the coupling.  Optionally, as mentioned in section \ref{subsec:hierarchy}, band gaps may also be used to enforce fixed magnetization dynamics with non-zero fermion Zeeman shifts.

In systematic studies of the cooling rates as a function of system parameters, for $\theta_{z-z'} \neq 0$ the rates typically were more sensitive than for $\theta_{z-z'} = 0$ to the other parameters, including to the released energy $E_{if}$. Looking for conclusions that are robust with chain length, we will focus on $\theta_{z-z'} = 0$. We find that in such a case where the spin chain is oriented along the magnetic field axis, the lattice transverse to $\vec B$ mostly contributes to enhancing magnetization-conserving couplings, while the longitudinal lattice mostly contributes in stabilizing the dipolar gas, but quickly becomes detrimental to the couplings. Deeper transverse lattices are preferable until one has to consider the deconfinement transition from 1D tubes or 1D Mott states to 3D BEC \cite{Ho2004, Stoferle2004, Cazalilla2006, Vogler2014}, and the quantum depletion that arises before it \cite{Xu2006}. 

Fully independent optimization of longitudinal (z') and transverse (x',y') boson lattice depths is possible using two laser sources, given the strong wavelength dependence of the boson light shift \footnote{Dynamics very similar to those presented below, section \ref{subsec:spinevolution_numericalresults}, can also be obtained using a single wavelength.}.
We set our parameters to $k_B T = 0.3 J$, a mean density of $\rho_0 = 3 \cdot 10^{19}$/m$^3$, i.e. a filling of 0.9 bosonic atoms per 3D site, in a transverse lattice depth $12 E_r^{Dy}$, and a longitudinal lattice depth $3.5 E_r^{Dy}$ (stabilization against dynamical instabilities operates from $\approx 2.5 E_r^{Dy}$). 
The first gap of the band structure is $h \times 5.2$ kHz wide. 

Based on 1D physics considerations neglecting the longitudinal lattice, the 1D Lieb-Liniger parameter $\gamma = m g_{1d} / \hbar^2 n_{1d} \approx 1.3$, while the transverse tunnelling is $t_{perp} \approx 6\cdot10^{-2} \mu_{1d}$, which according to \cite{Cazalilla2006} and consistently with the experiment \cite{Vogler2014} would ensure being deep in the 3D coherent regime. From the 3D lattice perspective, we have $U/(2 t_{x'} + 2 t_{y'} + 2 t_{z'}) \simeq 0.9$, far from the critical value $\simeq 5.8$ where the Mott transition could be reached. Following \cite{Xu2006}, we estimate quantum depletion based on Bogoliubov theory in the homogeneous system limit to be $N_{qd} / N_0 = 1/N_0 \int d^3q \, V /(2 \pi)^3 v(\vec q)^2 \simeq 5\%$. The transverse lattice depth is thus kept conservatively low to ensure the BEC 3D coherence.

\begin{figure}
  \begin{center}
      \includegraphics[width=0.7\columnwidth]{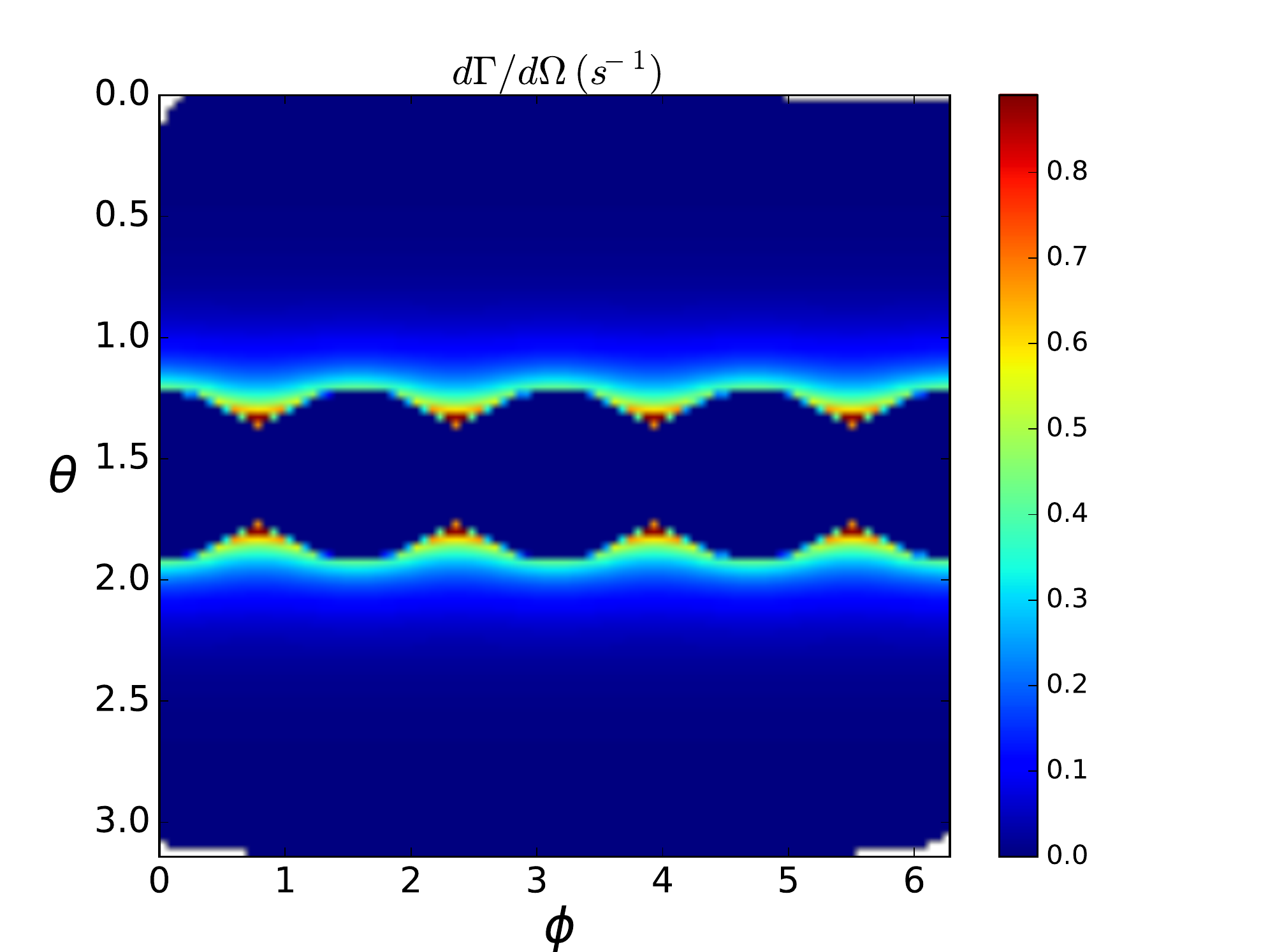}
      \caption{{\bf Phonon emission diagram (rate per solid angle), enhanced by the lattice,} of a length-two spin chain, from the triplet state $\ket{T_0}$ to the singlet state $\ket{S}$. Parameters are as in fig.\,\ref{fig:dispersion}b, and the magnetic field is oriented along the spin chain. 
      \label{fig:radiationdiagram_withlattice}}
  \end{center}
\end{figure}

In this setting, the surface of resonant phonon wavevectors is reduced to two sheets, orthogonal to $z'$ (figure \ref{fig:dispersion}). Due to the diverging density of states at the edges of the Brillouin zone, the phonon radiation pattern is heavily affected, as visible in fig. \ref{fig:radiationdiagram_withlattice}. As the lattice is more confining in the $(x', y')$ directions,  the terms of $\sum_{\vec n} [...]  {\rm{F[|w_B|^2]}} ( - \vec q -\vec n K  ) V_{col} (\vec q + \vec n K)$ in equation \ref{eq:developcoupling2_tightbinding} bear most weight for $\vec n$ in the $(x',y')$ plane, in which $V_{col}$ has constant sign. Thus, the summation is constructive.

Finally, we estimate with these settings that for the representative process $\ket{T_0} \rightarrow \ket{S}$ in a length-2 chain, the number of contributing phonon modes is, for $V = (20 \, \mu m)^3$, $N_{modes} \simeq 1.7$, and for $V = (100 \, \mu m)^3$, $N_{modes} \simeq 19$. The incoherent description is thus justified without the need to consider evaporation for large but realistic BECs.

\subsection{Spin chain evolution}
\label{subsec:spinevolution_numericalresults}
We now present the dissipative evolution of spin chains with the above parameters. To get a hint at how our simple picture scales towards the large system limit, we will repeat our calculation for varying 1D chain lengths $L$. Although our theory is valid for arbitrary chain length, we are limited by the computation time of the radiation pattern and anisotropic density of states, which have to be evaluated for all transition energies between spin chain eigenstates. The evolutions shown below rely on a three step approach. (i) Find all spin chain eigenstates, in a basis that also diagonalizes total pseudo-spin length and total spin projection along the polarizing field. (ii) Compute the rates between all of these states - this being the computationally limiting part as $L$ is increased. We use for efficiency a tight binding approximation in the $H_{int}$ matrix element along the two strong axes $(x',y')$ (see Appendix B), and along all axes for the $V_{\vec q}$ and thus the Bogoliubov description of the bath. 
(iii) Given all rates, simulate the dissipative evolution from a statistical distribution of the eigenstate probabilities.

\begin{figure}
  \begin{center}
      \includegraphics[width=0.7\columnwidth]{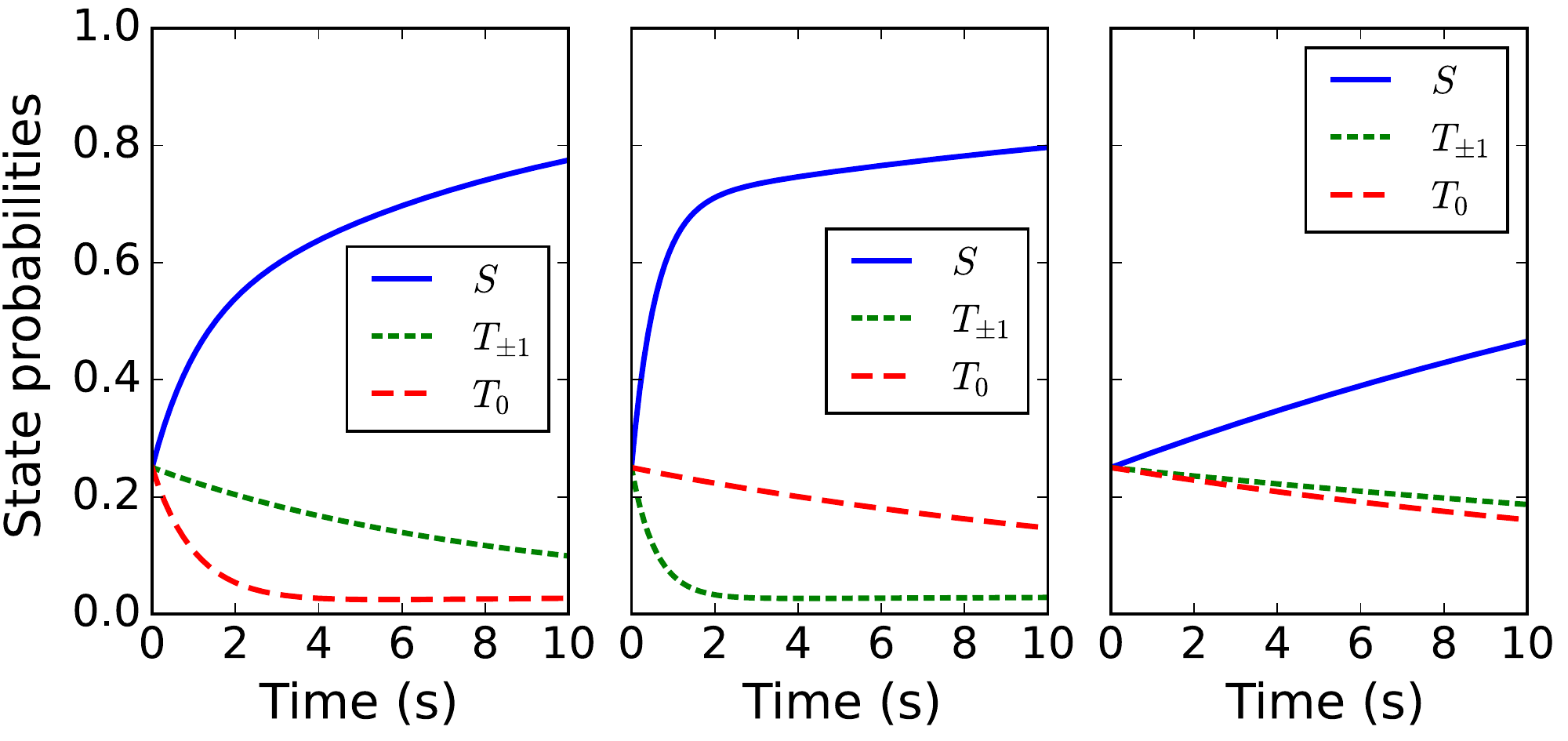}
      \caption{ {\bf Evolution of the eigenstate occupation probabilities} for a chain of length 2, at three different angles between magnetic field and spin chain axis: $\theta_{z-z'} = 0, \pi/4, \pi/2 $ from left to right. Parameters are as in fig.\,\ref{fig:dispersion}b. Here the system is prepared at free magnetization, such that all eigenstates are initially equally populated. The occupation probabilities of the eigenstates ($\ket{T_{\pm 1}}$) are always superimposed on this plot.
       \label{fig:evolutionofpops}}
  \end{center}
\end{figure}

First, we show in figure \ref{fig:evolutionofpops} the evolution with time of the probabilities of the four eigenstates from a length-2 spin chain, for various angles $\theta_{z-z'}$ of the chain axis with respect to $\vec B$. The cooling dynamics shows very strong sensitivity with the alignment, as a consequence of the anisotropy of the dipolar interaction. For parallel alignment ($\theta_{z-z'} = 0$), the fixed-magnetization transition $\ket{T_0} \rightarrow \ket{S}$ has the fastest rate, with timescale $\approx$\,0.9\,s. 
As will be demonstrated later, preparing the initial chain at total magnetization close to 0 (i.e., starting from a balanced spin mixture in the chain) may then provide the fastest dissipative route to low energy antiferromagnetic chains.

\begin{figure}
  \begin{center}
      \includegraphics[width=0.6\columnwidth]{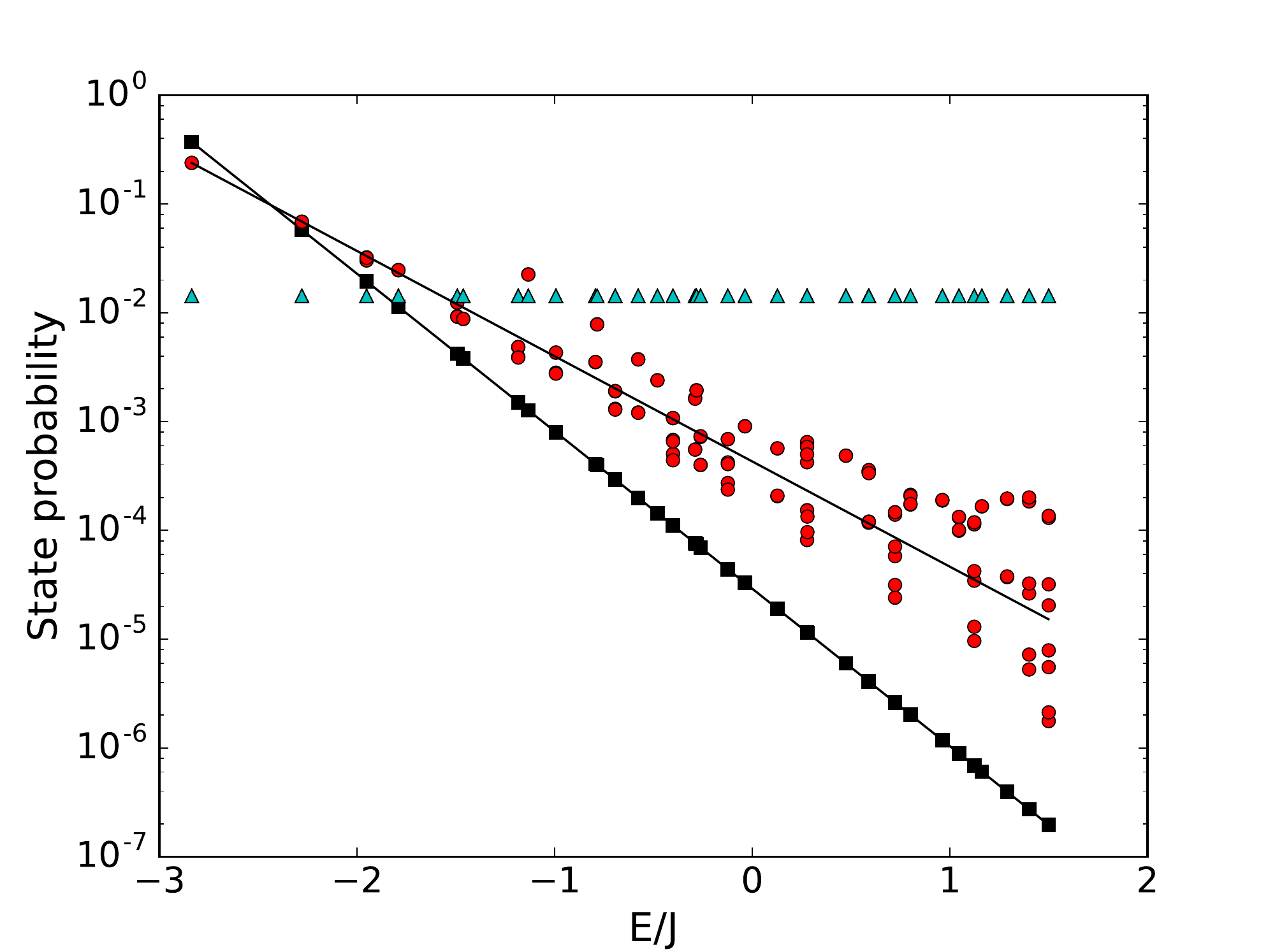}
      \caption{{\bf Convergence to the thermal spin state}, for 7 particles in the chain. Probabilities of the spin chain eigenstates, ordered by energy, at different times of the evolution. 
      Initially  (cyan triangles), all states with $\pm 1/2$ magnetization, and only those, are assumed equally populated. After full equilibration (black squares, 100\,s numerical evolution), the final probabilities of all the $2^7$ chain eigenstates reveal an exponential dependence as a function of energy. Two degenerate ground states have same probability. Red disks correspond to an intermediate equilibration time of 2s. Lines: fit with exponential functions, returning spin temperatures $k_B T_S \simeq 0.45 J$ after 2\,s, $k_B T_S = 0.3 J$ after 100\,s. Parameters as in fig.\,\ref{fig:radiationdiagram_withlattice}.
      \label{fig:thermalstate}}
  \end{center}
\end{figure}

We demonstrate in Fig.\,\ref{fig:thermalstate} that the spin chain reaches thermal equilibrium with the bath. The plot shows the occupation probabilities of the various spin chain eigenstates after several evolution times, for a chain of length L = 7. We start the evolution from a state with minimal magnetization, i.e. all states with total magnetization $\pm$1/2 have same probability, while all the other states have zero initial probability. This corresponds to a state of infinite spin temperature with the constraint of $\pm$1/2 magnetization.
We observe that the evolution converges towards a distribution of probabilities exponential with respect to energy, i.e., a thermal distribution in the many-body states \textit{at free magnetization}. We furthermore apply an exponential fit to the equilibrated distribution and extract a spin temperature of $k_B T_S = 0.3 J$ which corresponds to the bath temperature. Two seconds of evolution already lead close to a thermal state at spin temperature $k_B T_S \simeq 0.45 J$.

\begin{figure}
  \begin{center}
\includegraphics[width=0.7\columnwidth]{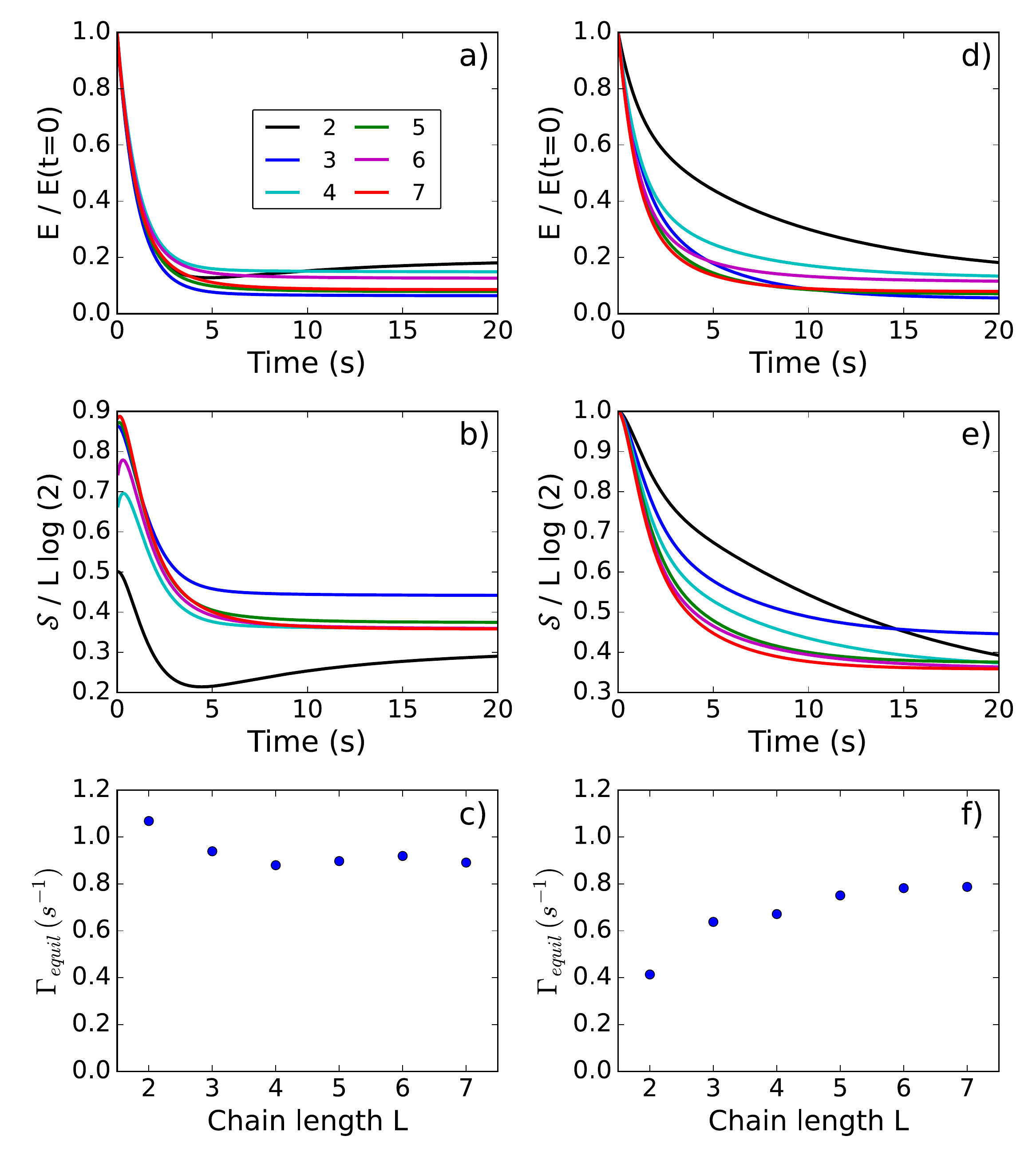}
      \caption{{\bf Evolution of global quantities for increasing spin chain length}. a-b-c: initial sample at minimal (0 or $\pm$1/2) magnetization. d-e-f: initial sample with free magnetization. The evolution is always freeing the magnetization. a-d: Spin chain energy as a function of time. b-e: Von Neumann entropy as a function of time. c-f: equilibration rate of the spin chain energy as a function of chain length. Parameters as in fig.\,\ref{fig:radiationdiagram_withlattice}.
       \label{fig:evolutionofthermo}}
  \end{center}
\end{figure}

To extend these observations to longer chains, we define global,  ``thermodynamic'' quantities describing the spin chain: the magnetic energy $E$, rescaled by its initial value, and the global von Neumann spin entropy $\mathcal{S} = - \sum_i p_i \log(p_i)$, where $p_i$ are the probabilities of the collective eigenstates. 
In fig.\,\ref{fig:evolutionofthermo}, we plot these quantities as a function of time, for chain lengths L ranging from 2 to 7. We actually compare the evolution from samples prepared at minimal initial magnetization, 0 or $\pm$1/2 depending on the chain length parity (left column), and prepared with free initial magnetization (right column). The dissipative evolution itself frees the magnetization, which is responsible for the initial entropy increase for samples prepared at fixed magnetization. 
Fig.\,\ref{fig:evolutionofthermo}(a-b) and (d-e) show that the evolutions of energy and entropy for increasingly large chains tend to converge with $L$, for both preparation protocols. 
The equilibration rates $\Gamma_{equil} = d/dt \ln\big(E - E(t \rightarrow \infty)\big)$ for both preparation protocols (fig.\,\ref{fig:evolutionofthermo}c,f), evaluated at early times, seem to quickly converge to similar and stable (size-independent) values. These observations suggest that long, mesoscopic chains may evolve with similar equilibration rates.

We observe in these plots, already after 2\,s, entropies per particle and spin temperatures that compare very favourably with the most recent experiments, e.g. \cite{Boll2016, Mazurenko2017}. 
Ultimately, provided the equilibration rate is sufficient to overcome imperfections such as light scattering, a key point is that the outcome is tied to the attainable BEC temperature; and indeed, as mentioned in section \ref{subsec:hierarchy}, extremely cold BECs can be produced experimentally, in bulk and in optical lattices\,\cite{Olf2015, Rey2006, Trotzky2010}. It thus decouples from the usual challenge to produce low entropy \textit{spinful} fermions in lattices, tackled e.g. by \cite{Mathy2012, Chiu2017}. 

\begin{figure}
  \begin{center}
      \includegraphics[width=0.5\columnwidth]{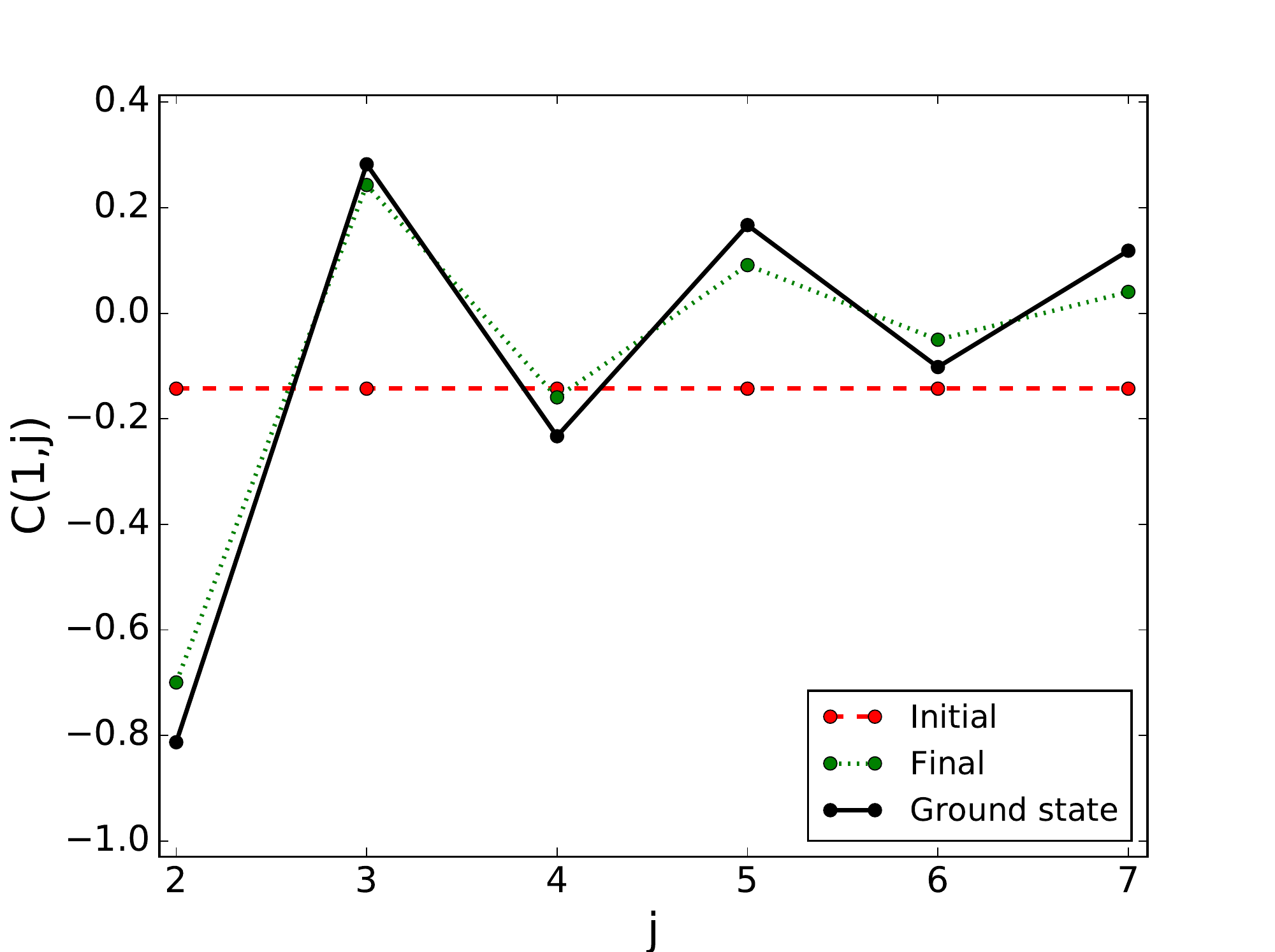}\\
      \caption{{\bf Spin correlators} for the 7-atom spin chain, initially (red, dashed line), after equilibration (green, dotted line), and for the two degenerate ground states (black, continuous line). Parameters as in fig.\,\ref{fig:radiationdiagram_withlattice}.  
       \label{fig:evolutionofcorrelators}}
  \end{center}
\end{figure}

To conclude, we show in Fig.\,\ref{fig:evolutionofcorrelators} spin correlations for the largest spin chain ($L=7$) before and after the evolution, and compare them to correlations at zero temperature. We use the following correlator definition, normalised to one for full correlation:
\begin{equation}
C(i,j) = 4 \langle \Sigma_{eff, i}^z \Sigma_{eff, j}^z \rangle 
\end{equation}
The flat negative correlation of the initial state is a finite size bias from the constrained initial magnetization $\pm 1/2$. For example, if the first site has magnetization $-1/2$, the $L-1$ other sites have a positive average magnetization per site $\propto 1/(L-1)$ that reflects upon the correlator. At the final temperature of $0.3 J$ we observe alternating spin correlations already similar to the ground state, spanning over the whole sample. This correlator is accessible on experiments, especially using quantum gas microscope techniques discriminating two spin states, as in \cite{Parsons2016, Preiss2015}. 
 
\section{Conclusion and possible extensions}

In this paper, we have discussed the possibility to couple the spin degrees of freedom of a spin chain in the Mott insulating regime to the motional temperature of a bath. We have in particular presented a realistic experimental configuration where the spin chain is made of a mixture of two spin states of an alkali atom trapped into an optical lattice, and the bath is provided by a spatially overlapping Bose-Einstein condensate of a strongly magnetic atom. The coupling between the collective spin states of the spin chain and the motional degrees of freedom of the BEC is provided by dipole-dipole interactions, which inherently provide the necessary spin-orbit-coupling. Simulating spin chains of length up to 7, and provided the mechanical temperature of the BEC is low enough,  we have found that low temperature ($T<J$) thermal states of the spin chain can be reached in a few seconds by the dissipative cooling associated to the dipolar interaction of the chain with the bath reservoir. 

We point out that the nature of the coupling between the spin chain and the BEC is collective. We find that the basic mechanism of cooling is a collective emission of a phonon in the BEC by the spin chain, which has direct analogies with cooperative effects in the context of spontaneous emission. 
This mechanism may also have interesting consequences for dipolar gas mixtures in new experiments \cite{Ilzhofer2018}, that could be apprehended with our framework.

The efficiency of cooling relies on the existence of phonon modes within the BEC whose energy is comparable to the energy to be dissipated within the spin chain (of the order of the super-exchange energy $J$), and whose momentum is comparable to the recoil momentum in the lattice $K/2$. These conditions imply rather strict experimental conditions, in particular on the lattice that controls the BEC phonon spectrum. 
To circumvent this difficulty, a natural extension would be to use a Fermi gas as a coolant, for which there exists a large density of available states sharing the same energy (close to the Fermi energy), but separated by the Fermi momentum (whose order of magnitude $k_F$ is typically comparable to $K/2$). This situation appears very promising for the release of low energy (i.e. comparable to $J$) spin-excitations from the spin chain to a Fermi gas.

In this work, we have considered explicitly the case of a coupling which frees the magnetization of the spin chain. To study antiferromagnetism, this implies that two single particle spin states of the species forming the spin chain should be kept almost degenerate, which can constitute an experimental difficulty. We have suggested that the use of low-enough magnetic field associated to micro-wave dressing is a possibility. However, we also pointed out that dissipative cooling \textit{at fixed magnetization} is possible as well, that circumvents such experimental complication. In this case, it is not necessary to artificially engineer near-degeneracy between two single-particle spin states within the chain. Suppression of magnetization-changing collisions is indeed ensured in a lattice due to energy conservation requirements, when the Zeeman energy falls in a gap of the Bloch band structure. 

Finally, we argue that spin-orbit coupling is an essential mechanism to \textit{generically} provide cooling of a spin-chain down to its lowest energy states. Indeed, in a given realization of an experiment, both the total magnetization and the collective spin of an assembly of spins are not necessarily those of its ground state. While this work focuses on spin-orbit coupling due to dipole-dipole interactions, we are also studying the possibility to artificially engineer spin-orbit coupling using Raman-dressed schemes.

\ack 

We acknowledge E. Mar\'echal and K. Kechadi for numerous discussions and critical reading of the manuscript. This work was supported by  Agence  Nationale de la Recherche (ANR Tremplin-ERC Highspin ANR-16-TERC-0015-01), the Conseil R\'egional d'Ile-de-France, Domaine d'Int\'er\^et Majeur Nano'K, Institut Francilien des Atomes Froids (projects MetroSpin and ACOST), and the Indo-French Center for the Promotion of Advanced Research (LORIC5404-1 and PPKC).

\section*{Appendix A. Controlling the quasi-degeneracy of two single-particle Zeeman states}
\setcounter{section}{1}

\label{subsec:quadraticshift}

Here, we estimate the possibility to employ micro-wave dressing in order to bring to degeneracy two adjacent Zeeman states of Potassium $^{40}K$ atoms in the lowest hyperfine manyfold F=9/2. We assume that the local magnetic field is 10 mG, which is a value that can be routinely produced in cold atom laboratories. For $^{40}K$, with Land\'e factor 2/9, the associated energy difference between two Zeeman states is $\Delta=$3\,kHz. We consider a micro-wave which is near resonant with the hyperfine structure of Potassium, at 1.285 GHz, with a detuning $\delta$. Following for example  \cite{Gerbier2006} we assume a $\pi$ polarized microwave field of Rabi frequency $\Omega/2 \pi = 100$\,kHz. Due to selection rules, the microwave dressing does not affect the $m_F$=-9/2 state, but introduces an energy shift $\Omega^2/4 \delta$ on the $m_F$=-7/2 state. In order to bring into degeneracy both states, we postulate $\Omega^2/4 \delta= -\Delta$, which sets the detuning, and the fraction of population off-resonantly excited to the upper hyperfine manyfold F'=7/2, $p=\Omega^2/4 \delta^2= 4 \Delta^2/\Omega^2$. For the experimental conditions which we chose, $p \simeq 0.004$. In case this value is too high ($i.e.$ in the case where inelastic collisions between Dy atoms and K atoms in F'=7/2 limit the lifetime), a higher Rabi frequency is needed. 

\begin{figure}
  \begin{center}
      \includegraphics[width=0.5\columnwidth]{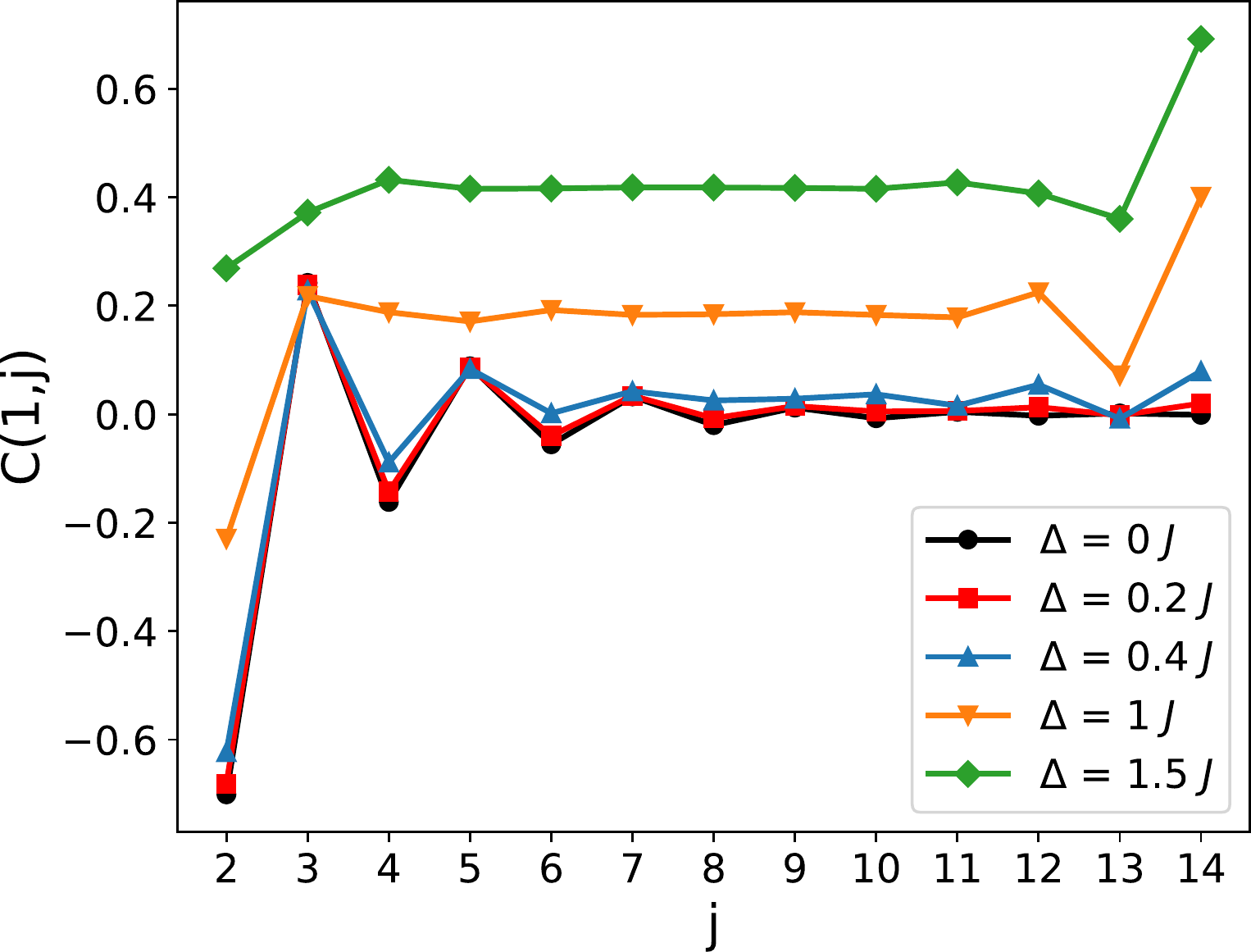}\\
      \caption{   {\bf Spin correlators for varying bias $\Delta/J$}. For each bias, the correlations of the thermal state at $k_B T = 0.3 \times J$ are shown, computed for a spin chain of length 14. The state is negligibly affected up to a bias of order 0.2$\times J$, above which the range of antiferromagnetic correlations decreases. Nearest-neighbor correlations remain antiferromagnetic around $\Delta = J$, and are destroyed slightly above.
       \label{fig:suppl:biascorrelations}}
  \end{center}
\end{figure}

This tuning actually offers the possibility to study the Heisenberg antiferromagnetic model with tunable external polarizing field. The required degree of tunability is to be significantly finer than the superexchange energy $J$: 
as shown in Fig.\,\ref{fig:suppl:biascorrelations}, the finite-range 1D antiferromagnetic correlations at our temperature are robust to a bias of  $0.2 \times J$. This bias corresponds to a magnetic field compensated to 0.4\,mG. The nearest neighbor antiferromagnetic correlation is robust to a bias of order $1 \times J$. 

\section*{Appendix B. Numerical considerations}
\setcounter{section}{1}

\subsection{Bloch states construction}
The Bloch states and their energies are computed by diagonalizing the non-interacting motion of the bosons and of the fermions in a truncated momentum space, ranging from $-30.5 K$ to $+30.5 K$, along each of the three directions of space. The Wannier functions constructed from this diagonalization are by default truncated in Fourier space to a support of $[- 4 K, 4K]$ along each dimension.

\subsection{Tight binding approximation}

We use twice in this paper a tight binding approximation, introducing the bosons Wannier functions and assuming ${\rm w_B}^{j,*}(\vec r) {\rm w_B}^{k}(\vec r) \equiv \delta^K(j,k) \lvert {\rm w_B}^j(\vec r) \rvert^2$.
When lowering the bath lattice depth along some direction of space, this becomes improper, such that the bath interaction with the spin chain, eq.\,\ref{eq:developcoupling2}, may be poorly approximated by eq.\,\ref{eq:developcoupling2_tightbinding}. In the conditions of section \ref{subsec:numerical}, we find that for a length-2 chain, the $\ket{T_0} \rightarrow \ket{S}$ transition rate differs between the two expressions by 12 $\%$. 
The agreement is improved to 1$\%$ if we apply the tight binding approximation only along the transverse axes $(x',y')$:
\begin{eqnarray}
\bra{f_{spin}; f_{bath}(\vec q)} H_{int}  \ket{i_{spin}; i_{bath}} = 
 \frac{\sqrt{N_0}}{V} \sum_{n_z\in \mathbb{Z}} \sum_{\vec m \in \mathbb{Z}^3} 
 \bra{f_{spin}} V_{col}( \vec q + \vec m K)\ket{i_{spin}}  
 \nonumber \\ 
 \Big( 
a^*_{n_z}[q_{z'}] a_{n_z- m_z}[0] u^*(\vec q)
 + a^*_{n_z}[0] a_{ n_z- m_z}[- q_{z'}] v^*(\vec q)
\Big) 
 \nonumber \\  {\rm{F[|w_F|^2]}} \big( \vec q + \vec m K \big) \cdot
{\rm{F[|w_B^{\perp}|^2]}} \big(-( q_{x'} + m_x K) \vec e_{x'} - (q_{y'} + m_y K) \vec e_{y'}\big) 
\end{eqnarray}
where we use a 3D Wannier wavefunction for the fermions ${\rm w_F}(x',y',z')$ but a 2D Wannier wavefunction for the bosons along the two tight axes, ${\rm w_B}^{\perp}(x',y')$.

Furthermore, to reduce computation time and study long chains, we calibrate on the length-2 chain calculation the acceptable truncation to Wannier functions in Fourier space, that reduces the reciprocal space summation $\sum_{n_z} \sum_{\vec m}$ while not affecting the rates within a 1$\%$ tolerance.

\subsection{Phonon direction sampling}
Each rate $\Gamma_{i \rightarrow f}$ requires to numerically sample the spherical coordinates $(\theta, \phi)$ of the emitted excitation wavevectors. We design an approximately homogeneous, isotropic sampling around any point on the sphere as follows.

i) The polar angle $\theta$ is varied by a fixed step : $\theta_n = \frac{\pi}{N_{\theta}} (n+ \frac{1}{2})$, with $n \in [0, N_\theta-1]$.

ii) All perimeters on the unit sphere, defined by a polar angle $\theta_n$,  are sampled with same angular spacing. This means that the azimuthal angle $\phi$ sampling depends on $\theta_n$. We furthermore sample each perimeter from a different, random starting point. This sums up as, for each  $\theta_n$, an azimuthal sampling $\phi_k = {\rm random}[0,2 \pi] + \frac{2 \pi}{N_{\phi}(\theta_n)} k$, with $N_{\phi}(\theta_n) = {\rm Integer} \Big( N_{\phi}^0  \sin(\theta_n)\Big) $ and  $k \in [0, N_\phi(\theta_n) -1]$. 

We observe that approximately isotropic sampling around any point on the sphere is obtained for $N_{\phi}^0 \simeq 2 N_{\theta}$. Then, we find that the cooling rates have converged to a $6 \cdot 10^{-2}$ stability using $N_\theta = 50$. 
Figures 4 to 6 are constructed with $N_\theta$ varying from 50 to 200, figures 7 to 10 with $N_\theta = 50$.

\section*{References}

%\bibliography{strontium} 

\end{document}